\documentclass{article}
\usepackage[top=.9in, bottom = .9in, left = 1in, right = 1in]{geometry}
\usepackage{amsmath}
\usepackage{graphicx}
\usepackage[font={small}, labelfont=bf]{caption}

\title{Chapter 4: Thermal Transport Measurement Techniques for Nanowires and Nanotubes}
\author{annie}
\date{\today}

\begin{document}
\vspace*{-1.5cm}
\noindent{\small This article should be cited as:\\A. Weathers, L. Shi, "\emph{Thermal Transport Measurement Techniques for Nanowires and Nanotubes}," Annual Review of Heat Transfer, Vol. 16, pp. 101-134, ed. G. Chen, DOI: 10.1615/AnnualRevHeatTransfer.v16.40, Begell House (2013)}
\vspace*{.7 cm}

\noindent {\large \textbf{Thermal Transport Measurement Techniques for Nanowires and Nanotubes}}\\

\noindent {\textit{Annie Weathers and Li Shi}$^{*}$}\\

\noindent {Department of Mechanical Engineering, The University of Texas at Austin, Austin, TX 78712, USA}\\

\noindent $^{*}$Address all correspondence to Li Shi E-mail: lishi@mail.utexas.edu \\

\begin{minipage}[adjusting]{16cm}
{\em Recent advances in the synthesis of inorganic and organic nanowires and nanotubes have
provided both components for various functional devices and platforms for the study of low-
dimensional transport phenomena. However, tremendous challenges have remained not only
for the integration of these building blocks into functional devices, but also in the characterization of the fundamental transport properties in these nanoscale model systems. In particular,
thermal and thermoelectric transport measurements can be considerably more complicated
than electron transport measurements, especially for individual nanostructures. During the
past decade, a number of experimental methods for measuring the thermal and thermoelectric properties of individual nanowires and nanotubes have been devised to address these
challenges, some of which are reviewed and analyzed in this chapter. Although the Seebeck
coefficient and electrical conductivity can also be obtained from some of the measurement
methods, this chapter is focused on measurement techniques of the thermal conductivity and
thermal diffusivity of nanowires and nanotubes. It is suggested that the limitations in the current experimental capability will provide abundant opportunities for innovative approaches to
probing fundamental thermal and thermoelectric transport phenomena in individual nanostructures.
}
\end{minipage}
\section{Introduction}

\indent Nanowire (NW) and nanotube structures have been explored as building blocks for electronic, photonic, thermoelectric, and thermal management devices.$^{1-3}$ For example, “bottom-up” synthesized semiconducting NWs and carbon nanotubes have been employed as
the conducting channel in field-effect transistors (FETs).$^{4-7}$ Meanwhile, the nanoscale
channel in “top-down” patterned Si FinFETs is essentially a NW.$^{8,9}$ In nanoelectronic devices, the high and nonuniform heat dissipation density and the resulting local hotspots are
detrimental to device performance and reliability, and present a major challenge. Phonon
scattering by interface roughness can considerably reduce the effective thermal conductivity of nanostructures including NWs, and is one of the causes of the local hotspots in
nanoelectronic devices.$^{10,11}$

However, the reduced thermal conductivity in NWs and other nanostructured materials is desirable for thermal insulation and thermoelectric materials. Thermoelectric materials are characterized by their dimensionless figure of merit, $ZT = S^2 \sigma T/\kappa$, where $S$ is the Seebeck coefficient, $\sigma$ is the electrical conductivity, $T$ is the absolute temperature, and $\kappa$ is the thermal conductivity that consists of a lattice or phonon contribution
($\kappa_L$) and an electronic contribution ($\kappa_e$). The best thermoelectric materials will therefore
have low thermal conductivity, and high electrical conductivity and Seebeck coefficient.
However, engineering such a material is nontrivial due to the substantial coupling between
these three parameters. For example, increasing the charge carrier concentration via doping can be used to increase the electrical conductivity, but often decreases the Seebeck
coefficient, and increases the electronic contribution to the thermal conductivity. Consequently, the highest ZT values reported for bulk single crystals have been limited to be
close to unity. In recent years, there have been a number of investigations of nanostructuring to enhance thermoelectric performance. The suppressed lattice thermal conductivity
in nanostructures can help to increase the $ZT$, when the power factor ($S^2\sigma$) is not suppressed as much as the lattice thermal conductivity.$^{12}$ This is possible when the interface
or boundary scattering mean free path in the nanostructures is shorter than the phonon
mean free path in the bulk crystal but longer or comparable to the bulk electron mean free
path. In this case, it is possible for the phonon mean free path and thermal conductivity
to be reduced more than the electron mean free path and electrical conductivity, respectively.

In addition to the classical effects of interface scattering, there have also been a number of theoretical studies of the quantum size effects on the thermoelectric power factor
($S^2\sigma$). As one of the first theoretical studies, Hicks and Dresselhaus$^{13}$ examined thermoelectric transport in 1D $\textrm{Bi}_2\textrm{Te}_3$ quantum wires, for which the wire diameter is comparable to
or smaller than the de Broglie wavelength of electrons. The electronic density of states in
such quantum wires can be highly asymmetric around the Fermi energy, which can result
in an enhanced power factor.$^{14,15}$

Similarly, as the diameter of the NW is reduced further below the dominant wavelength of thermal phonons, ranging from on the order of 1 nm at room temperature to tens of nanometers at low temperatures, there exist only a few 1D phonon sub-bands with
well separated wave vector components along the radial direction of the NW. Because
of the requirement of energy and momentum conservation in phonon-phonon scattering,
some of the phonon-phonon scattering events allowable in bulk crystals are eliminated in
quasi-1D NWs because of the modified phonon dispersion. In conjunction with an atomically smooth surface, the suppressed phonon-phonon scattering may reverse the dependence of the lattice thermal conductivity on the diameter of nanowires, and can lead to
increasing thermal conductivity with decreasing diameter, when the diameter is reduced
below a threshold value on the order of the phonon wavelength.$^{16}$ Carbon nanotubes
and polymer fiber chains are two representative examples of such quasi-1D systems of
potentially high thermal conductivity that is desirable for thermal management applications.

Therefore, recent advances in the synthesis of inorganic and organic NWs and nanotubes have provided both components for various functional devices and a platform for
the study of low-dimensional transport phenomena. However, tremendous challenges have
remained not only for the integration of these building blocks into functional devices, but
also in the characterization of the fundamental transport properties in these nanoscale
model systems. In particular, thermal and thermoelectric transport measurements can be
considerably more complicated than electron transport measurements, even in bulk samples. For example, issues such as radiation loss, heat loss to the thermometers, and contact
thermal resistance can lead to large uncertainty in thermal conductivity measurements.
To emphasize these complications, Tye$^{17}$ wrote in the preface of a 1969 text on thermal conductivity that ``the situation has not been helped when poor experimental work
had led to suggestions that new transport mechanisms exist, only for them to be eliminated by a later more careful experimental investigation.'' If issues already existed in
the thermal measurements of bulk size samples, these issues can become pronounced in
nanostructure samples of a much smaller dimension, especially individual NWs and nanotubes that are difficult to handle and require miniaturized thermometers in the measurements.

During the past decade, a number of experimental methods for measuring the thermal and thermoelectric properties of individual NWs and nanotubes have been devised to
address these challenges. Some of these methods are reviewed and analyzed in this chapter. As summarized in Table 1, these methods include those based on NW samples with a
steady state temperature difference applied to the two ends, such as measurements based
on suspended resistance thermometer microdevices and bimaterial cantilever thermal sensors. In addition, several techniques based on electrical self-heating and optical heating of
the NW samples have been reported, including the celebrated 3$\omega$ method and other steady
state or pulsed heating techniques. Besides electrical resistance thermometry, optical non-contact thermometry techniques including micro-Raman spectroscopy and time domain
thermoreflectance (TDTR) measurements have been explored for thermal measurements of
individual NWs, nanotubes, and NW arrays. Although the Seebeck coefficient and electrical conductivity can also be obtained from some of the measurement methods, this chapter
is focused on measurements of the thermal conductivity and thermal diffusivity for NWs
and nanotubes.

\begin{table}[h]
\begin{tabular}{|c|c|c|c|c|c|c|}
\hline
\textbf{Techniques}                                                                & \textbf{Heating Method}                                                                & \textbf{Sensing Method} & $\kappa$ & $\alpha$ & $C_p$ & $\sigma, S$ \\ \hline
Suspended mesa structure                                                           & \begin{tabular}[c]{@{}c@{}}Steady state \\ end heating\end{tabular}                    & Noise thermometry       & x &   &    &      \\ \hline
\begin{tabular}[c]{@{}c@{}}Suspended resistance \\ thermometer device\end{tabular} & \begin{tabular}[c]{@{}c@{}}Steady state \\ end heating\end{tabular}                    & Resistance thermometry  & x &   &    & x    \\ \hline
T-junction sensor                                                                  & \begin{tabular}[c]{@{}c@{}}Steady state\\ end heating\end{tabular}                     & Resistance thermometry  & x &   &    &      \\ \hline
\begin{tabular}[c]{@{}c@{}}Bimaterial \\ cantilever sensor\end{tabular}            & \begin{tabular}[c]{@{}c@{}}Steady state \\ end heating\end{tabular}                    & Cantilever deflection   & x &   &    &      \\ \hline
\begin{tabular}[c]{@{}c@{}}Cantilever resistance \\ thermometry\end{tabular}       & \begin{tabular}[c]{@{}c@{}}Steady state \\ end heating\end{tabular}                    & Resistance thermometry  & x & x &    &      \\ \hline
3$\omega$                                                                            & \begin{tabular}[c]{@{}c@{}}Modulated self-electrical \\ heating\end{tabular}           & Resistance thermometry  & x & x & x  & x    \\ \hline
\begin{tabular}[c]{@{}c@{}}Transient electrothermal \\ techniques\end{tabular}     & \begin{tabular}[c]{@{}c@{}}Pulsed electrical or \\ optical heating\end{tabular}        & Resistance thermometry  & x & x &    &      \\ \hline
\begin{tabular}[c]{@{}c@{}}Micro-Raman \\ spectroscopy\end{tabular}                & \begin{tabular}[c]{@{}c@{}}Optical absorption or \\ self-electrical heating\end{tabular} & Raman peak shift        & x &   &    &      \\ \hline
\begin{tabular}[c]{@{}c@{}}Time domain \\ thermoreflectance\end{tabular}           & Optical heating                                                                        & Thermoreflectance       & x & x &    &      \\ \hline
\end{tabular}
\caption{Summary of thermal measurement techniques for nanowires and nanotubes.
$\kappa$, $\alpha$, $C_p$ , $\sigma$, and $S$ refer to the thermal conductivity, thermal diffusivity, specific heat,
electrical conductivity, and Seebeck coefficient of the sample. The 'x' indicates the property
that was obtained by each technique.
}
\end{table}

\section{Suspended Mesa Structures for Thermal Conductance Measurements of Suspended Beams}
One of the first thermal measurements of nanostructures was reported by Tighe et al.,$^{18}$
who developed a method of measuring the thermal conductance of patterned, suspended
GaAs nanobeams in the temperature range of 1.5 - 6 K. The measurement device was patterned from a wafer composed of three layers, i.e., a topmost conducting layer of heavily
doped GaAs from which a serpentine heater and electrodes were patterned, a middle undoped GaAs layer from which the central thermal reservoir and GaAs nanobeams were defined, and finally a sacrificial AlAs layer. Figures 1(a) and 1(b) show an image of the measurement device, which has the four supporting GaAs nanobeams with the cross-sectional
dimension of $200 \times 300$ nm. The thermal conductance was obtained by supplying a direct
current (DC) heating to one of the serpentines on the central membrane and monitoring
the membrane temperature rise with a small modulated sensing current through the other
serpentine resistance thermometer. The thermal conductance, $G = Q/\Delta T$, was obtained
from the measured DC heating rate, $Q$, and the temperature rise of the heating membrane
determined from the sensing voltage.

\begin{figure}[h!]
\includegraphics[width = 1\textwidth]{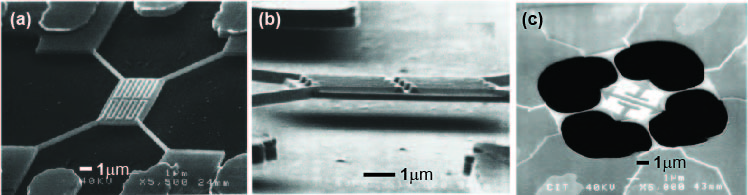}
\caption{(a,b) SEM images of the measurement device from Tighe et al.$^{18}$ with four patterned GaAs nanobeams supporting a central GaAs membrane with serpentine resistance
heater and thermometer patterned on top. (c) SEM image of the device of Schwab et al.$^{19}$
used to measure the quantum of thermal conductance, with a central SiN$_x$ membrane supported by four SiN$_x$ beams with Nb leads on top. (a,b) Reprinted with permission from
Tighe et al.$^{18}$ Copyright 1997, American Institute of Physics. (c) Reproduced from Schwab
et al.$^{19}$ with permission from Macmillan Publishers Ltd: Nature. Copyright 2000.
}
\end{figure}

A similar device was used to measure the quantum of thermal conductance of $\textrm{SiN}_x$
nanoconstrictions by Schwab et al.$^{19}$ Their device consisted of a $\textrm{SiN}_x$ central membrane
with four supporting $\textrm{SiN}_x$ beams, a patterned Cr/Au serpentine heater and sensor, and
four Nb leads, as shown in Fig. 1(c). The $\textrm{SiN}_x$ nanoconstrictions were patterned with a
$\cosh^2{x/\lambda}$ geometry, where $x$ is the lateral coordinate along the lead and $\lambda$ was designed
to be 1 $\mu$m. Such a geometry was intended to achieve a transmission coefficient close to
unity for long-wavelength phonons, which dominate thermal transport at temperatures of
$<1$ K. In addition, at temperatures of $<9.2$ K, the Nb leads become superconducting and
therefore generate no electrical heating in the supporting beams. The electron temperature
of the Au film was obtained by measuring its Johnson noise at a minimum power with
the use of a sensitive superconducting quantum interference device (SQUID).$^{20}$ Below
a cutoff temperature of $\sim$800 mK, when only the four lowest-lying modes are excited,
they found a plateau in the thermal conductance corresponding to $16g_0$, where $g_0$ is the
quantum of thermal conductance, $\pi^2 k_B^2 (T/3h)$,$^{21}$ and the factor of 16 arises because of
the four $\textrm{SiN}_x$ phonon waveguides each with four 1D branches, including one longitudinal,
two transverse, and one torsional branch. At this cutoff temperature, the dominate phonon
wavelength is comparable to the dimensions of the nanoconstrictions, which is 200 nm at
its smallest dimension.

\section{Suspended Resistane Thermometer Microdevices for Thermal and Thermoelectric Measurements of Suspended Nanowires and Nanotubes}
Shi,$^{22}$ Kim et al.,$^{23}$ and Shi et al.$^{24}$ fabricated a suspended platinum resistance thermometer (PRT) device for measuring the thermal conductivity of individual carbon nanotubes.
The device was later used by Li et al.$^{25}$ for the thermal measurement of Si NWs and by
Mavrokefalos et al.$^{26}$ for the measurement of InAs thin films. The device consists of two
adjacent thermally isolated $\textrm{SiN}_x$ membranes supported by long $\textrm{SiN}_x$ beams, as shown
in Figs. 2(a) and 2(b). A Pt thin film serpentine is patterned on each $\textrm{SiN}_x$ membrane, and
was connected to four Pt leads allowing for four-probe measurement of the resistance of the
serpentine. In recent designs, two electrodes were patterned on each membrane for measurements of the Seebeck coefficient and the electrical conductivity of the samples.$^{26,27}$ The membranes and $\textrm{SiN}_x$ beams were suspended above a through-substrate hole, which
allowed for transmission electron microscopy (TEM) characterization of the nanostructure
sample assembled on the membrane.

\begin{figure}[h!]
\centering
\includegraphics[width = .7\textwidth]{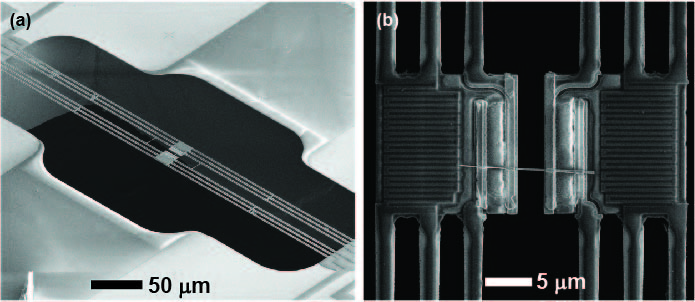}
\caption{SEM images of (a) the suspended device used for thermal and thermoelectric measurements of individual NWs and (b) an InAs NW with Ni/Pd evaporated on the four contacts through a SiN$_x$ shadow mask.
}
\end{figure}

\subsection{Sample Preparation}

Several methods have been used to place a NW across the two suspended membranes of
the resistance thermometer device. One method involves dispersing the NWs in a solvent
and drop casting the suspension onto a wafer containing many of the suspended devices.
Occasionally, one NW is trapped between the two membranes of a suspended device. This
method requires a NW suspension with sufficiently high NW density in order to achieve
a good assembly yield. In another method, a tungsten probe attached to a micromanipulator can be used to manually pick up the NW and place it across the device. This may
be achieved either with a nanomanipulator inside the vacuum chamber of a scanning electron microscope (SEM), or simply with a home-built micromanipulator stage underneath
an optical microscope. The process is clean and leaves no solvent residue on the NW surface. However, this method is not effective for some fragile NW materials. In addition, the
adhesion between the NW and $\textrm{SiN}_x$/Pt membrane is weak compared to the sample prepared with the drop-casting method, where residue or moisture left by the solvent helps to
enhance the adhesion between the nanostructure and the two membranes.$^{28}$

Another challenge in the sample preparation procedure is making electrical contact between the NW and the Pt electrodes patterned at the edge of each $\textrm{SiN}_x$ membrane. The
presence of native oxide on the surface of a NW can prevent direct electric contact between the assembled NW and the underlying Pt electrodes on the two $\textrm{SiN}_x$ membranes.
Mavrokefalos et al.$^{29}$ have shown that annealing in a forming gas containing 5\% hydrogen in nitrogen was able to reduce the surface oxide of a $\textrm{Bi}_2\textrm{Te}_3$ NWs assembled on the
suspended device so as to obtain ohmic electrical contact between the NW and the Pt electrodes without any deposited metal. However, this method has not been effective for other
NW materials. Focused ion beam (FIB)-assisted deposition and electron beam-induced
deposition (EBID) of Pt/C can be used to make electrical contact and improve the thermal contact to a variety of suspended NWs, as shown by a number of works.$^{27,30-34}$ However,
it has been shown that there is considerable spreading of the Pt-C within a 5 $\mu$m radius
of the electron beam spot during EBID.$^{35}$ To reduce the influence of metallic contamination of the nanowire surface, Tang et al.$^{36}$ have evaporated Ni through the windows of a
SiNx shadow mask directly onto the two contacts between a porous Si film and the two
suspended membranes. Because of the presence of native oxide on the Si film, the evaporated metal did not make electrical contact to the sample. In another work, Weathers et
al.$^{37}$ succeeded in depositing metal contacts through a shadow mask to make electrical
contact to an InAs NW assembled on the suspended device shown in Fig. 2(b), and etched
in $\textrm{BCl}_3$ plasma before deposition. Further efforts along this direction can potentially lead
to reliable, clean contact between different NWs and the suspended devices.

\subsection{Two-Probe Thermal Measurement Procedure}
The thermal conductance of a suspended NW can be obtained by electrically heating one
PRT and monitoring the temperature rise of the two $\textrm{SiN}_x$ membranes. During the thermal
measurement, the sample is placed in a variable-temperature cryostat. Heat transfer to the
surrounding gas molecules is minimized by evacuating the sample space of the cryostat to a
vacuum level better than $10^{-5}$ torr with the use of a turbomolecular pump. When the internal thermal resistance of each membrane is much smaller than the thermal resistance of the
supporting beams and that of the sample, the temperature on each membrane is uniform, as
verified by numerical heat transfer analysis.$^{38,39}$ The radiation loss from the circumference
\begin{figure}[h]
\centering
 \includegraphics[width = .8\textwidth]{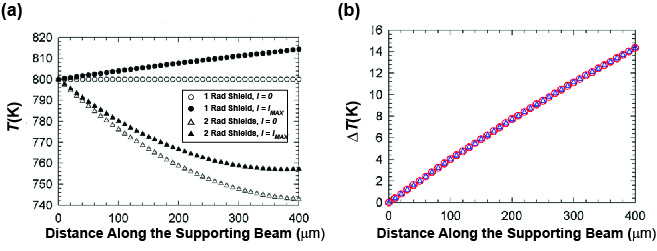}
 \caption{(a) Temperature profile along the supporting SiN$_x$ beams with and without electrical heating and with one and two radiation shields, for $T_0$ = 800 K. (b) The difference in
temperature $\Delta T = T (I) - T (I = 0)$ for one (triangles) and two (circles) radiation shields.
The additional shield in the two-shield configuration is thermally anchored to the sample
stage. Compared to the nonlinearity in the temperature profile for the absolute temperature,
the difference in temperature is essentially linear. Reproduced from Moore and Shi$^{39}$ with
permission from IOP Publishing. Copyright 2011.
}
\end{figure}
of the long supporting beams can be accounted for in the heat diffusion equation given by
\begin{equation}
\frac{d^2T}{dx^2} + \frac{\dot{q}}{\kappa_b} = \frac{\sigma \epsilon P (T^4 - T_0^4)}{\kappa_b A_b}
\end{equation}
where $x$ is measured from the junction between the substrate and the $\textrm{SiN}_x$ beam, $\kappa_b$, $L_b$, $A_b$, $P$, and $\dot{q}$ are the thermal conductivity, length, cross section, perimeter, and the volumetric heating rate of the supporting beam, respectively, $\epsilon$ and $\sigma$ are the emissivity of
the sample and the Stefan-Boltzmann constant, and $T_0$ is the temperature of the environment enclosing the sample. The ratio between the radiative and conductive heat transfer
through the supporting beam, $Q_{rad}/Q_{cond}$, can be found by multiplying the right-hand side
of Eq. (1) by $L^2_b/\Delta T_{x=0\rightarrow L}$, where $\Delta T_{x=0\rightarrow L}$ is the temperature difference between the
two ends of the beam. Because of the large thermal resistance ($L_b/\kappa_b A_b$) of the supporting
beams for each membrane, on the order of $10^7$ K/W, this term is not negligible when $T_0$
differs significantly from the average temperature of the beam. The effects of radiation loss
can be clearly seen in Fig. 3(a), calculated by Moore et al.$^{39}$ by finite element modeling,
which shows that the use of one radiation shield yields a highly nonlinear temperature profile of the beam supporting the sensing membrane at a high $T_0 = 800$ K. As one important
consequence, the sensing membrane temperature $T_s$ can be almost 60 K lower than $T_0$
when no electrical heating ($I = 0$) is supplied to the heating membrane.

However, the nonlinearity in the temperature profile can be overcome by considering
instead the change in temperature between nonzero heating current, $I$, and zero heating
current, $I = 0$. Subtracting Eq. (1) for $I = 0$ from the same equation for $I \neq 0$ gives
\begin{equation}
\frac{d^2 \Delta T}{dx^2} + \frac{\dot{q}}{\kappa_b} = \frac{\sigma \epsilon P}{\kappa_b A_b} \left (T(I)^4 - T(I = 0)^4 \right )
\end{equation}
where $\Delta T (x) \equiv T(x, I)-T (x, I = 0)$. During measurements, $\Delta T$ is usually kept at $<10$ K
for $T_0 > 300$ K and at $<5$ K for lower temperatures. Consequently, the $T(x, I)^4 - T(x, I =
0)^4$ term is considerably smaller than $T(x, I)^4 - T_0^4$ for the case of inadequate radiation
shielding. As a result, the $\Delta T(x)$ profile obtained from the numerical analysis of Moore
et al.$^{39}$ with radiation loss taken into account is nearly linear, even for the case of only one
radiation shield and the sample stage temperature at $\Delta T_0 = 800$ K, as shown in Fig. 3(b).

With the additional radiation loss term in the right-hand side of Eq. (2) ignored, a current $I$ to the heating membrane results in additional heat conduction $\Delta Q = \kappa_b A_b \Delta T(L)/L_b$
to the substrate from each supporting beam without the heating current, and $\Delta Q = \kappa_b
Ab \Delta T(L)/L_b + Q_l /2$ for each beam carrying a DC heating current, where $Q_l = I^2 R_l$,
and $R_l$ is the electrical resistance of the Pt lead on the SiN$_x$ beam. Because the temperature is uniform on each of the two membranes, $T(x = L)$ is equal to $T_h$ and $T_s$ for
the supporting beams of the heating and sensing membranes, respectively. With the additional radiation loss ignored, $Q_h + 2Q_l = 6(\kappa_b A_b /L_b )(\Delta T_h + \Delta T_s ) + Q_l$, where
$\Delta T_h \equiv T_h(I) - T_h(I = 0)$ and $\Delta T_s \equiv T_s(I) - T_s (I = 0)$. Therefore, the thermal
resistance of the six supporting beams for each membrane is
\begin{equation}
R_b = \frac{\Delta T_h + \Delta T_s}{Q_h + Q_l}
\end{equation}
The ratio of radiation loss from the suspended NW to the heat conduction through the
NW is proportional to $L^2 /d$, where $L$ and $d$ are the suspended length and diameter of the suspended NW, respectively. When $L$ is on the order of $10 \,\mu$m or shorter, radiation loss is
negligible compared to heat conduction through most NW materials even at temperatures
as high as $800$ K, as shown in Ref. 39.

In addition to radiation loss, there also exists a thermal contact resistance between the
NW and the SiN$_x$ membrane. A resistance circuit including this contact resistance is presented in Fig. 4 together with a characteristic temperature profile along the nanowire. The
two-probe thermal resistance of the sample can then be obtained from
\begin{equation}
R_{total} = R_S + R_{c1} + R_{c2} = R_b \frac{\Delta T_h - \Delta T_s}{\Delta T_s}
\end{equation}
where $R_S$ refers to the intrinsic thermal resistance of the sample, and $R_{c1}$ and $R_{c2}$ are
the contact thermal resistances to the two ends of the NW. The temperature rise of the
membranes, $\Delta T_h$ and $\Delta T_s$, are obtained by measuring the electrical resistance of the two
PRTs. At different heating current ($I$) supplied to the heating PRT, the four-probe electrical resistance ($R_s$) of the sensing PRT can be measured with a sinusoidal current from a
\begin{figure}[h]
 \centering
 \includegraphics[width = .5\textwidth]{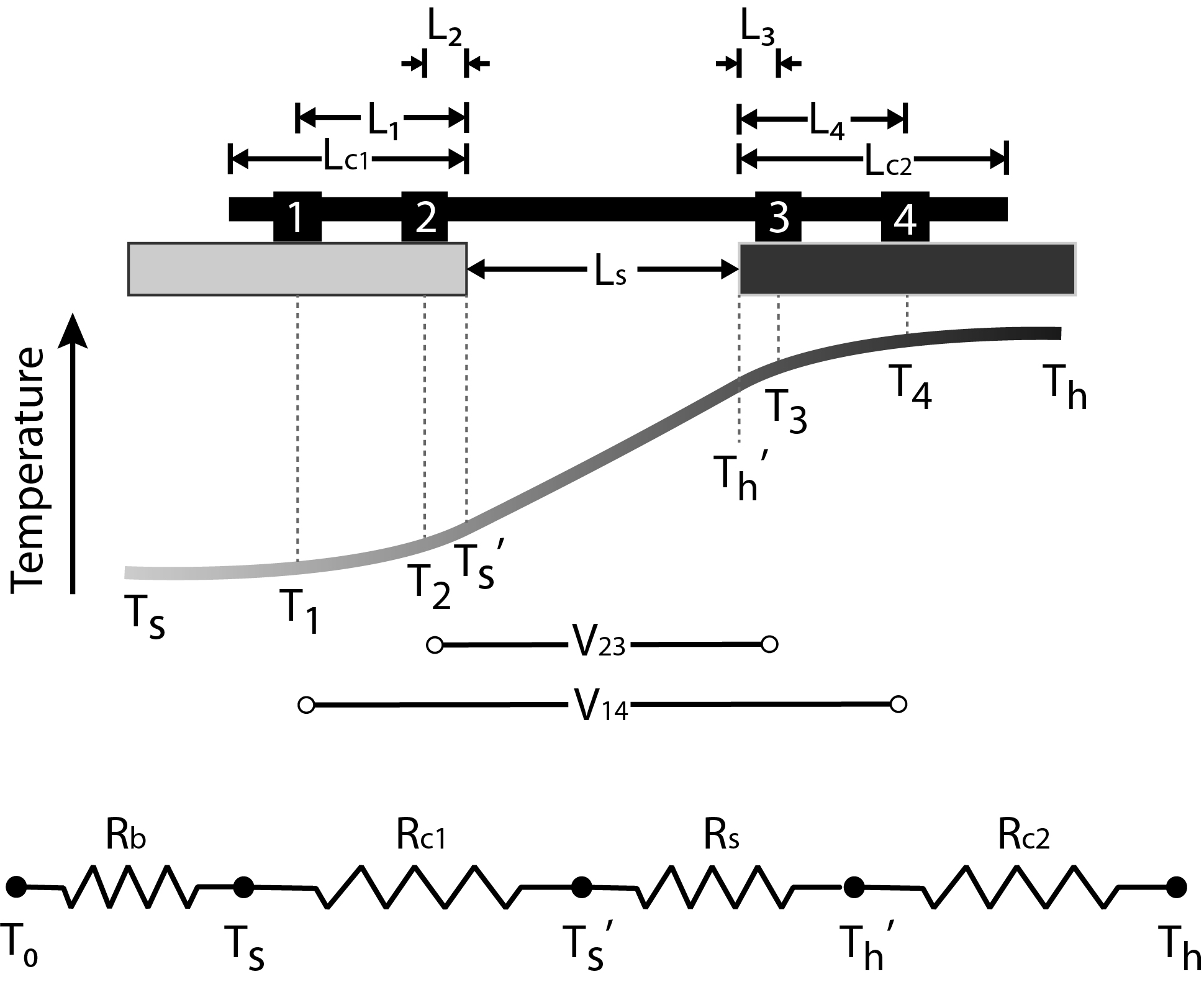}
 \caption{Four-probe thermoelectric measurement schematic with definitions of the length,
temperatures, and thermal resistances used in the analysis.
}
\end{figure}
lock-in amplifier. The temperature rise of the sensing membrane is obtained as
\begin{equation}
\Delta T_s (I) \equiv \frac{\Delta R_s}{dR_s(I=0)/dT}
\end{equation}
where $\Delta R_s = R_s (I)- R_s (I = 0)$, and the temperature coefficient of resistance, $dR_s (I = 0)/dT$, of the PRT must be determined accurately from the measured $R_s (I = 0)$ versus
$T_s (I = 0)$ curve. As discussed above, with inadequate radiation shielding or at high or
low temperatures, $T_s (I = 0)$ deviates considerably from the sample stage temperature $T_0$,
which is measured with a temperature sensor attached to the sample stage. Thus, although
Eqs. (3) and (4) are still accurate for the case of inadequate radiation shielding, proper radiation shielding is necessary to reduce the temperature difference between $T_s (I = 0)$ and
$T_0$ during the temperature calibration of the sensing PRT. The use of one additional radiation shield thermally anchored to the sample stage can considerably reduce the temperature
difference between the sensing PRT and the shield, as shown in Fig. 3(a).
\begin{figure}
 \centering
 \includegraphics[width = .6\textwidth]{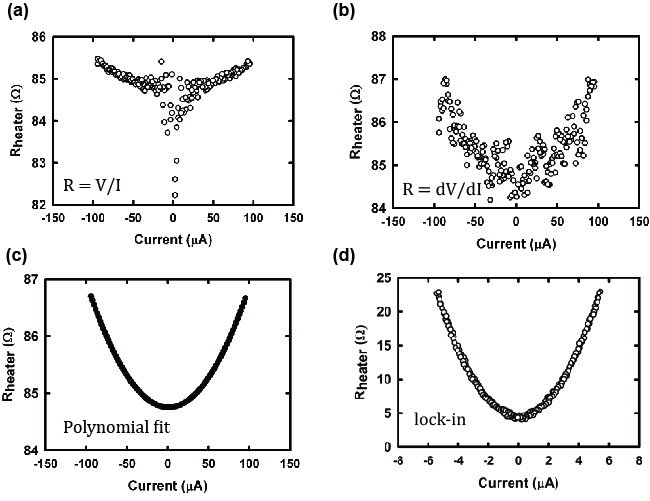}
 \caption{The resistance of the heating PRT as a function of the applied DC current determined by three methods: (a) the voltage rise, $V$, divided by the applied current, $I$; (b) the
five-point local derivative of $dV /dI$; (c) the analytic derivative to the third-order polynomial
fit of $V$ versus $I$; (d) the resistance of a different PRT measured with a lock-in amplifier by
coupling a small AC current with a DC heating current.
}
\end{figure}

The resistance of the heating PRT can be obtained from the measured $I - V$
curve as $R_h = V /I$. However, the obtained $R_h$ becomes noisy when $I$ approaches zero, as
shown in Fig. 5(a). To address this issue, the differential resistance of the heating PRT can
be calculated from the local slope of the $I - V$ curve as $R_h^{'} = dV /dI$ [Fig. 5(b)]. However,
the obtained $R_h^{'} (I)$ and $R_h (I)$ are much noisier than the $R_s (I)$ measured with the use of
a lock-in amplifier. It is possible to smooth the data by taking an analytic derivative of the
third-order polynomial fit to the measured $\Delta V$ versus $I$ [Fig. 5(c)], however, this provides
no improvement in the actual measurement uncertainty. This problem can be overcome
by coupling a small sinusoidal current $i_{1\omega} e^{i\omega\tau}$ to the large DC heating current $I$. The
first-harmonic component of the voltage drop, $v_{1\omega} e^{i\omega\tau}$, across the heating serpentine can
be measured with a lock-in amplifier, and used to obtain the AC resistance of the heating
serpentine as $R_h = v_{1\omega} /i_{1\omega}$ . In this case, the electrical heating in the heating serpentine
becomes
\begin{equation}
Q_h = \left (I^2 + 2Ii_{1\omega}e^{i\omega\tau} + i_{1\omega}^2 e^{i2\omega\tau} \right ) R_h(I)
\end{equation}
Consequently, $\Delta T_h (I)$ contains a steady state component, a $1\omega$ component, and a higher-order term. The higher-order term is negligible compared to the steady state and $1\omega$
components, because of the relatively small applied sinusoidal current, giving $\Delta T_h (I) = a_0 I^2 + 2a_1 Ii_{1\omega} e^{i\omega\tau}$. If $\omega$ is small compared to $1/\tau$, where $\tau$ is the thermal time constant
of the device, $\Delta T_h (I)$ responds fully to the $1\omega$ heating term just like it does to DC heating
(i.e., $a_0 = a_1$). In the high-frequency limit, $\omega$ is much larger than $1/\tau$ so that $\Delta T_h (I)$ does not respond to the modulated heating (i.e., $a_1 \ll a_0$).
The resulting temperature rise in the heating PRT for the two limiting cases is
\begin{subequations}
\begin{align}
\Delta T_h(I) \equiv \frac{\Delta R_h}{3dR_h/dT} \quad &\textrm{for} \quad\omega \ll 1/\tau \\
\Delta T_h(I) \equiv \frac{\Delta R_h}{dR_h/dT} \quad &\textrm{for} \quad\omega \gg 1/\tau
\end{align}
\end{subequations}
The factor of three difference is verified by the measured $\Delta R_h$ versus $\omega$ relation shown
in Fig. 6. A detailed transfer function analysis of this $1\omega$ measurement technique was
given by Dames and Chen.$^{40}$ Because of the relatively large thermal time constant of the
thermally isolated measurement device, the high-frequency limit can be readily achieved
by using a frequency of $> 1000$ Hz which is well below the frequency range where electrical
capacitive coupling can cause a noticeable phase lag between the measured $v_{1\omega}$ and $i_{1\omega}$.
\begin{figure}[h]
 \centering
 \includegraphics[width = .5\textwidth]{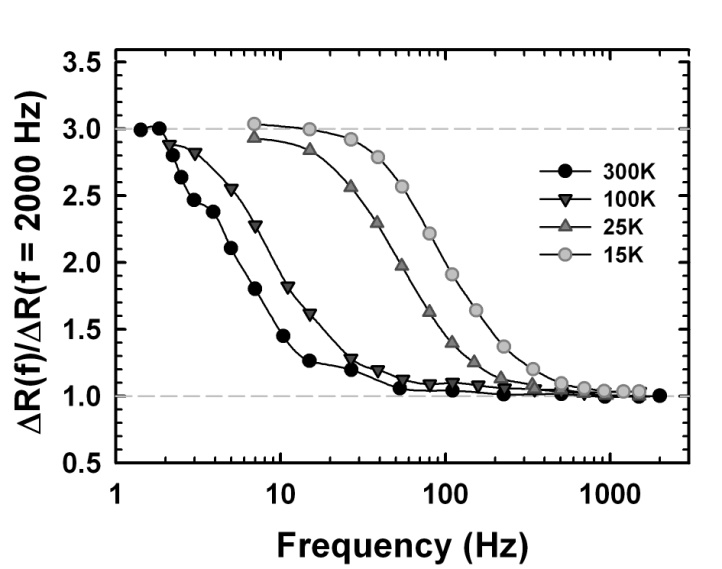}
 \caption{The measured resistance of the PRT as a function of the lock-in amplifier frequency,
which shows a factor of three difference between the low and high frequency limits. Reproduced from Shi et al.$^{24}$ 
with permission from ASME. Copyright 2003.
}
\end{figure}

Returning to the problem of accurately determining the temperature coefficient of resistance, Figure 7 shows the measured $R_s (I = 0)$ versus $T_0$ curve and the $dR_s/dT$ of a PRT
in the temperature range between $4$ and $700$ K with two radiation shields. The resistivity
is roughly linear at high temperatures, and falls approximately as $T^5$ at low temperatures.
Depending on the nature of the impurities in the PRT, the resistance may increase again at
very low temperature; the measured resistance in Fig. 7(a) clearly shows such dependence at $<50$ K. The temperature coefficient of resistance, $dR/dT$, may be found from the local derivative to the $R$ versus $T$ curve, or alternatively, by fitting the measured resistance
versus temperature to a linear fit or to the Bloch-Gruneisen (BG), $T^5$ law,
\begin{equation}
R(T) = R_0 + A\left (\frac{T}{\theta_D}\right ) \int\limits_0^{\theta_D} \frac{x^5 \,\textrm{d}x}{\left (e^x - 1 \right ) \left (1- e^{-x} \right )}
\end{equation}
where $R_0$ is the residual resistance, $\theta_D$ is the Debye temperature, and $A$ is a constant.
Wingert et al.$^{41}$ have used the BG relationship to fit the measured resistance of a PRT
in the temperature range of $60-380$ K with the fitting parameters $A$, $R_0$, and $\theta_D$, and
found that a linear approximation to the resistance can lead to an $8$\% difference in the
measured thermal conductivity at temperatures on the order of $100$ K due to a peak in
the temperature derivative at low temperature. Although the BG fit to the data shows a
considerable improvement over a constant $dR/dT$ value, the BG formula makes several
key assumptions, which include neglecting the interactions of electrons with higher energy
phonons and the assumption that electron states with the same wave vectors have nearly
the same velocity. However, as the Fermi surface becomes increasingly complex, interband
scattering becomes important. In addition, when the temperature is sufficiently high, the
resistance can deviate from Eq. (8).$^{42}$ In addition, Eq. (8) neglects any contribution from
the Kondo effect, and thus cannot explain the resistivity below 10 K for certain materials
with strong magnetic moments or a high concentration of magnetic impurities.$^{43}$ In this
case, a polynomial fit to the data or the determination of the local derivative can be most
accurate. Figure 7(b) shows the difference in the calculated $dR/dT$ using a linear fit, a
BG fit, and a local three-point derivative. For the linear fit and BG fit, only the data greater
than $70$ K is considered. Both the linear fit and BG fit fail at temperatures of $<100$ K for
this particular PRT. A more reasonable fit to the BG formula is possible by considering all
data points, however, this gives unrealistic values for the fitting parameters, and therefore
provides no clear advantage over using a simple polynomial fit.

\begin{figure}[h]
 \centering
 \includegraphics[width = .8\textwidth]{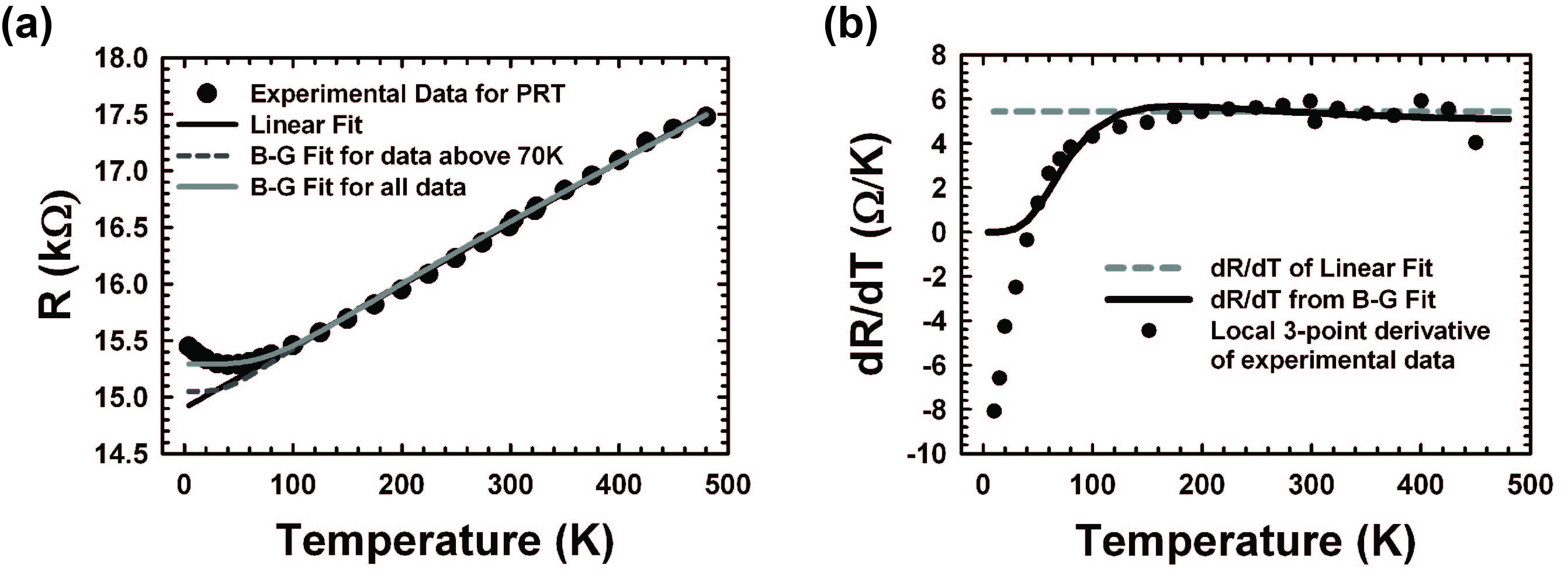}
 \caption{(a) Measured resistance (circles) as a function of temperature for a PRT of a suspended resistance thermometry device, 
 with a linear and BG fit. (b) The temperature coefficient of 
resistance calculated with a local numerical derivative (circles) and the analytic
derivative to the BG fit and linear fit.
}
\end{figure}
\subsection{Measurement Sensitivity}

Figure 8 shows the electronic connections for measuring the resistance change of both
PRTs with the use of two lock-in amplifiers. The noise in the measured resistance of the
two PRTs was found to be dominated by a $\sim 40 \times 10^{−3}$ K random fluctuation in the sensing
membrane temperature when $R_0$ is near 300 K.$^{24}$ Because the thermal conductance of the
supporting beam, $G_b \equiv 1/R_b$, is about $1 \times 10^{−7}$ W/K and $\Delta T_h$ should be maintained at
$<10$ K for $T_0$ near 300 K, $\Delta T_s$ approaches the order of the random temperature fluctuation
when the sample thermal conductance, $1/R_{total}$, is less than $0.5 \times 10^{−9}$ W/K, which is
equivalent to the thermal conductance of a $20$ nm-diameter, 3 $\mu$m-long NW with a low
thermal conductivity of $5$ W/m K. For measuring this and other NWs of a smaller $G_{total}$,
a large $\Delta T_h$ is required unless the random noise can be reduced. A recent work$^{41}$ has
reported the use of a Wheatstone bridge circuit to enhance the measurement sensitivity. In
that work, the sensing PRT was connected in parallel with an additional serpentine PRT on
an adjacent device on the same chip. The resistance ($R_{s,p}$) of this reference PRT is very
close to that of the sensing PRT ($R_s$). In addition, two other resistors are paired with the
measurement device outside the cryostat, including a high-precision resistor $R_0$ and
a potentiometer $R_p$. The bridge is initially balanced by zeroing the measured differential
voltage $V_D$ by adjusting $R_p$. During the measurement, DC Joule heating of one membrane
results in a temperature rise of the sensing membrane and resulting increase in $R_s$. With
an AC voltage $V_0$ applied to the bridge circuit, the bridged voltage $V_D$ is measured with
a lock-in amplifier. The resistance of the sensing membrane and two leads is then found
from
\begin{equation}
R_s = R_0 \left (\frac{V_D}{V_0} + \frac{R_p}{R_p + R_{s,p}} \right )^{-1} - R_0
\end{equation}
The bridge circuit takes advantage of the fact that fluctuations in the substrate temperature
will affect both $R_s$ and $R_{s,p}$ equally, generating a net zero change in the bridge voltage
$V_D$. The sample conductance can then be found in a similar manner as described previously
from the sensing and heating membrane resistance. Because the effect of the substrate temperature fluctuation in the sensing PRT is canceled by a similar fluctuation in the reference
PRT, the process increases the $G_{total}$ sensitivity to roughly $1 \times 10^{−11}$ W/K, when a large
voltage is applied to the bridge circuit and the measurement time constant is long. The
bridge circuit also yields an additional benefit because only the change in the voltage drop
along the sensing PRT is measured by the lock-in amplifier, so that the resolution of the
lock-in amplifier can be fully utilized.

However, even without a NW bridging the two membranes, a background thermal conductance ($G_{bg}$) as high as $0.3 \times 10^{−9}$ W/K at 300 K can be measured between the two
membranes,$^{30}$ because of residual gas molecules, thermal radiation, and heating of the
substrate. In the case that $G_{total}$ is comparable to or smaller than $G_{bg}$, the measurement uncertainty can be appreciable. Hence, this method cannot be used to measure a sample with
$G_{total}$ much smaller than about $0.3 \times 10^{−10}$ W/K.
\begin{figure}[h]
 \centering
 \includegraphics[trim = 0mm 0mm 1000px 0mm, clip, width = .7\textwidth]{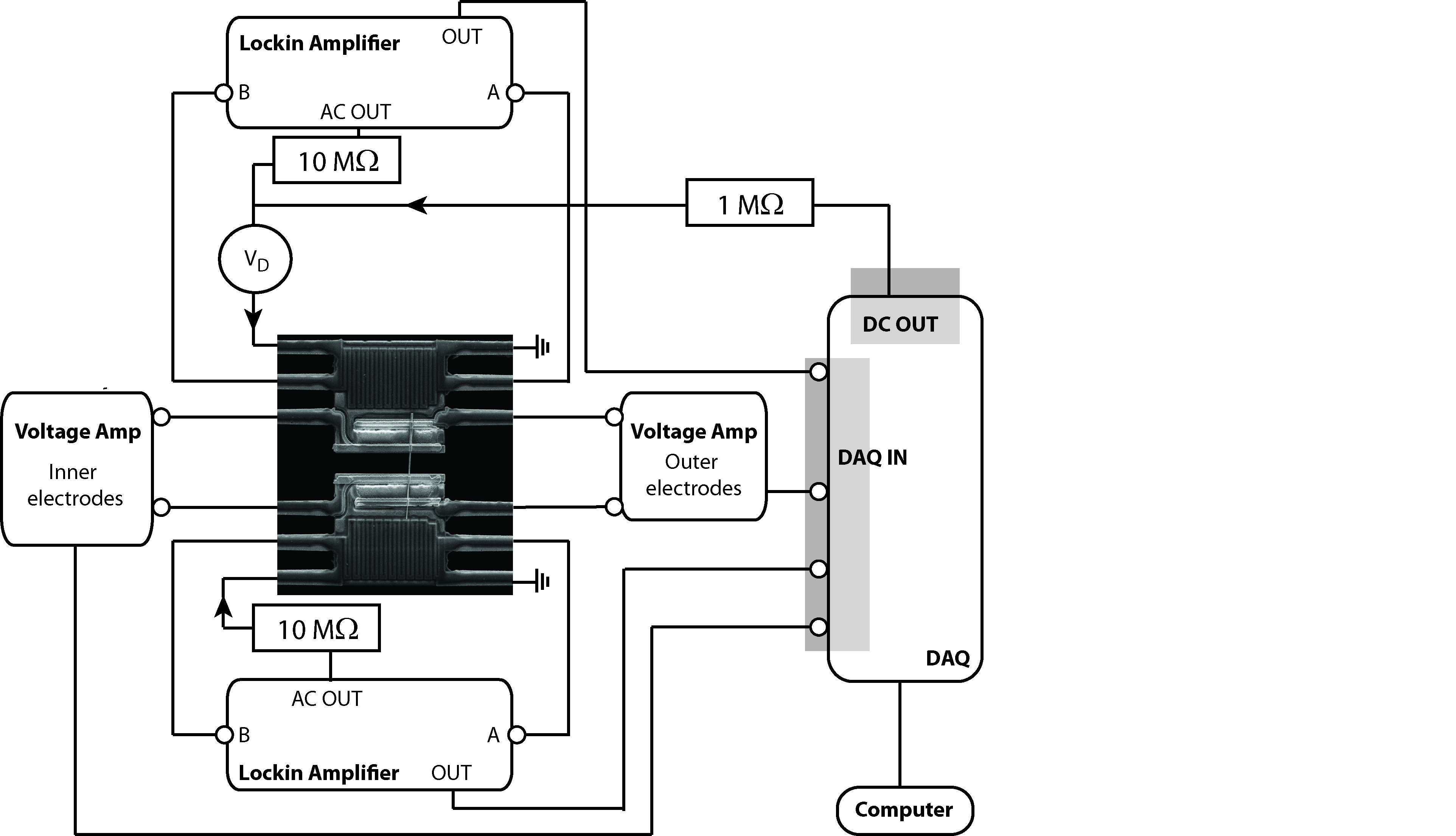}
 \caption{Schematic of the measurement setup for four-probe thermoelectric property measurement of suspended nanostructures.
}
\end{figure}

\subsection{Contact Resistance and Four-Probe Thermoelectric Measurement Procedure}

According to the thermal resistance circuit in Fig. 4, the as-measured two-probe thermal
resistance of the sample includes a contribution from the contact resistance between the
NW and the two membranes. In an attempt to minimize the contact thermal resistance,
Hippalgaokar et al.$^{44}$ have patterned an array of Si NWs from a thin Si membrane, with
the PRT/SiN$_x$ membrane deposited directly onto the supporting substrate on either ends of
the nanowires. By measuring NW arrays of a similar cross section and different lengths,
they determined that the intrinsic thermal resistance of the suspended NW array was much
larger than the constriction thermal resistance at the nanowires ends, and the interface
thermal resistance between the SiNx membrane and the Si pad. Length-dependent thermal
resistance has also been employed to determine the contact thermal resistance between
carbon nanotubes and the suspended membranes.$^{45}$ In comparison, for a NW assembled by
drop casting between two suspended membranes, the contact thermal resistance between
the NW and the membrane can vary greatly depending on the material, contact area, and
the interface adhesion energy.

To resolve the issue of the unknown contact resistance, Mavrokefalos et al.$^{26}$ have devised a four-probe thermoelectric measurement procedure to obtain the contact thermal
resistances at the two ends of a NW sample and thus determine the intrinsic thermal conductivity, Seebeck coefficient, and electrical conductivity of the suspended NW. In
this method, two thermovoltages were measured between the two outer electrical contacts
and the two inner electrical contacts to the NW, $V_{14} = S(T_1 - T_4)$ and $V_{23} = S(T_2 - T_3)$
(Fig. 4), where $S$ is the Seebeck coefficient of the NW and is assumed to be uniform along
the entire NW and much larger than that of the metal electrodes.

Figure 4 also shows a representative temperature distribution across the NW, considering the nanowires to be a fin of constant cross section. The temperature across the suspended segment is linear when the radiation loss from the NW is negligible. Because of
heat transfer through the NW-membrane contact, the temperature distribution along the
supported NW varies approximately exponentially. The thermal contact resistance can then
be calculated from the fin resistance formula,
\begin{equation}
R_{c,i} = \frac{1}{\kappa A m \tanh{m L_{c,i}}} \quad i = 1,2
\end{equation}
where $\kappa$ is the thermal conductivity, $A$ is the cross-sectional area of the suspended NW,
$L_{c,i}$ is the length of the NW in contact with the membrane as defined in Fig. 4, and
$m = \sqrt{hb/\kappa A}$, where $h$ is the thermal contact conductance per unit area and $b$ is the contact width between the NW and membrane. From the thermal resistance circuit in Fig. 4,
together with the definition of the sample resistance, $R_S = L_s/\kappa A$, and the fin temperature profile, the dimensionless temperature is related to the two measured thermovoltages according to
\begin{equation}
\gamma \equiv \frac{\gamma_{14}}{\gamma_{23}} = \frac{T_1 - T_4}{T_2 - T_3} = \frac{V_{14}}{V_{23}}
\end{equation}
where
\begin{align}
\begin{split}
\gamma_{ij} \equiv \frac{T_i - T_j}{T^{'}_h - T^{'}_s} = 1 &+ \frac{1 - \cosh{m(L_{c,1}-L_i)}/\cosh{mL_{c,1}}}{L_S m \tanh{mL_{c,1}}}\\
&+ \frac{1 - \cosh{m(L_{c,2}-L_i)}/\cosh{mL_{c,2}}}{L_S m \tanh{mL_{c,2}}}
\end{split}
\end{align}
The intrinsic thermal resistance and Seebeck coefficient of the NW can be found from
\begin{align}
S = \frac{\alpha V_{23}}{\gamma_{23}(T_h - T_s)}; \alpha &= \frac{T_h - T_s}{T_h^{'} - T^{'}_s} = 1+ \frac{1}{L_s m} \left [ \frac{1}{\tanh{m L_{c,1}}} + \frac{1}{\tanh{m L_{c,2}}} \right ]\\
R_s &= \frac{R_{total}}{\alpha}
\end{align}
For CrSi$_2$ NWs, Zhou et al.$^{27}$ have found the contact resistance accounts for ∼10\% of the
measured $R_{total}$. In comparison, Mavrokefalos et al.$^{26}$ have found this contribution to be
as high as 15-20\% for InAs thin films.

In addition to the thermal conductivity and Seebeck coefficient, the four-probe electrical conductivity can also be measured on the suspended resistance thermometer device,
allowing for characterization of all three axial thermoelectric properties on the same NW.
The electrical current used for the four-probe $I - V$ measurement of the NW can result
in Peltier cooling and heating at the two contacts, and result in a temperature difference
between the two membranes. The temperature difference can introduce a thermovoltage
component in the measured four-probe voltage. This thermovoltage component can be determined from the previously obtained $S$ of the NW as well as the temperatures of the two
membranes measured with the use of the two PRTs.

This four-probe thermoelectric measurement method is limited to electrically conducting NW or nanofilm samples with a sufficiently large Seebeck coefficient that is uniform
along the entire length. However, for atomically thin NWs, nanotubes, and films such as
graphene, the Seebeck coefficient depends sensitively on the surface charges, and may
be rather different between the suspended segment and supported segment of the sample.
Hence, this method has not been used to obtain the contact thermal resistance to graphene
or carbon nanotubes.

\section{T-Junction Sensor for Thermal Conductance Measurement}
Fujii et al.$^{46}$ reported a T-junction nanofilm sensor for measuring the thermal conductivity
of a multiwalled carbon nanotubes (CNTs) inside an SEM or TEM column, which can be
used to characterize the crystal structure and cross section of the measured nanostructure
sample. In this measurement, one end of the CNT was mounted, with the assistance of
focused electron beam irradiation, to a heat sink at one end and to the center of a patterned
suspended Pt nanofilm sensor at the other. Figure 9 shows a variation of this measurement
setup reported by Dames et al.,$^{47}$ in which a scanning tunneling microscope (STM) tip
was used as the heat sink on one end of the CNT, which could be manipulated to bring the
CNT into contact with the electrically heated nanofilm sensor. The average temperature
rise of the electrically heated nanofilm sensor was determined from the measured electrical
resistance before and after the CNT made contact. For the case when a CNT with thermal
resistance $R_{CNT}$ is brought into contact with the midpoint, $x = 0$, of the hotwire sensor,
the average temperature rise of the sensor can be found from the heat conduction equation
as
\begin{equation}
\theta = \frac{1}{12} Q R_{HW} \left [1 - \frac{3}{4} (1 + \gamma^{-1})^{-1} \right ]
\end{equation}
where $Q$ is the Joule heat generation in the hotwire, $\gamma = R_{HW} /4R_{CNT} , R_{HW} = L_{HW}/\kappa_{HW} A_{HW}$, and $L_{HW}$, $A_{HW}$, and $\kappa_{HW}$ are the length, cross section, and thermal conductivity of the hotwire. The derivative of Eq. (15) with respect to $\gamma$ reaches a maximum when
$\gamma$ approaches unity, and zero at sufficiently low and high values of $\gamma$. Hence, the maximum
measurement sensitivity is obtained when $R_{HW}$ is designed to be close to $4R_{CNT}$. In the
measurement, $R_{HW}$ can be obtained based on Eq. (15) from a separate measurement made
before the CNT made contact to the hotwire, where $\gamma = 0$.

This analysis assumes that the CNT makes contact at exactly the midpoint of the sensor.
In the case that the sample is offset slightly from $x = 0$, a correction factor is needed.
Dames et al.$^{47}$ have shown that this fractional error is approximately $8(l/L_{HW})^2$ and is
$<10$\% as long as the offset, $\ell$, is $<0.112$ times the length of the hotwire. In addition,
radiation loss is not accounted for in this conduction analysis. For both the hotwire and
the nanotube, the ratio of radiative heat transfer to conductive heat transfer scales as the
product of the fin parameter ($\beta$) and length, $\beta L = h_{rad}P/\kappa A L$, where $h_{rad}$ is the heat
transfer coefficient for radiation and is approximately equal to $h_{rad} = 4\epsilon_{rad}\sigma T^3$ , and $\epsilon$,
$\kappa$, $A$, and $P$ are the emissivity, thermal conductivity, cross section, and perimeter of the
nanotube or hotwire. Dames et al.$^{47}$ derived the relative error caused by neglecting the
\begin{figure}[h]
 \centering
 \includegraphics[ width = .8\textwidth]{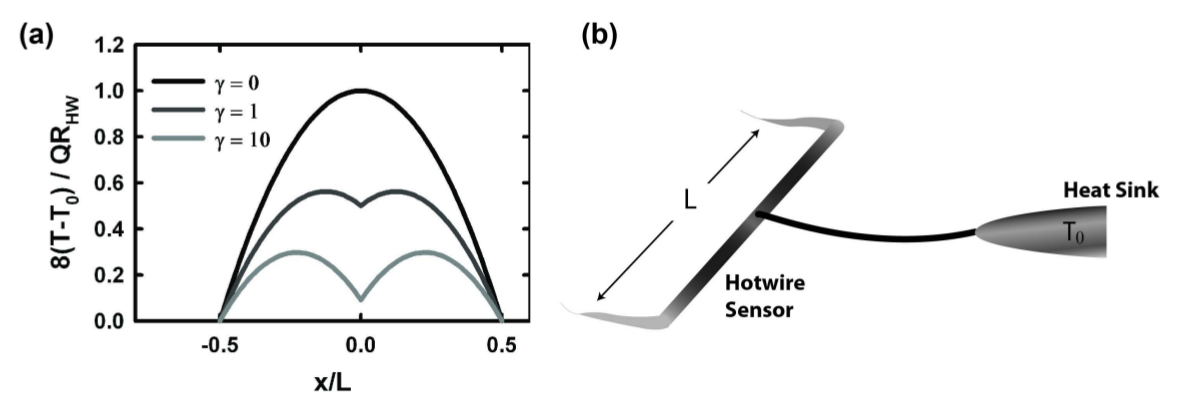}
 \caption{(a) Temperature profile along the length of the hotwire probe from Dames et al.$^{47}$
for varying ratios, $\gamma$, of the hotwire and sample thermal resistance. (b) A schematic of the
measurement setup for the same work. Reproduced with permission from Dames et al.$^{47}$
by American Institute of Physics. Copyright (2007).
}
\end{figure}
radiation loss to be approximately $(\beta L)^2/10$. This relative error can increase from $<0.48$\%
for a 3 $\mu$m-diameter, 2 mm-long Pt hotwire, to up to 32\% for a 0.5 $\mu$m-diameter, 5 mm-
long Pt hotwire. Although the radiation loss from the hotwire can be reduced by increasing
its diameter and thermal conductivity and reducing its length, doing so would reduce $\gamma$, and hence the sensitivity in the measurement of $R_{CNT}$.

\section{Direct Thermal Conductance Measurement with Bimateral Cantilever Sensor}
Recently, Shen et al.$^{48}$ have reported a method based on a Si$_3$N$_4$/Au bimaterial cantilever
thermal sensor to measure the thermal conductivity of stretched polyethylene nanofibers.
In their measurement, the polyethylene nanofibers were drawn directly from a polymer gel
with the use of the bimaterial cantilever, with a draw ratio between 60 and 800. The fiber
was cut at a distance of $\sim 300$ $\mu$m from the cantilever. The cut end of the fiber was then
attached to a microthermocouple that could be externally heated. Because of the thermal
expansion mismatch between the two constituent materials of the cantilever, the heat flow
($Q_C$) from the fiber into the cantilever tip can be obtained from the cantilever tip deflection,
which is determined from the position of the reflection of a laser beam focused on the
cantilever.

The thermal measurement was conducted in vacuum to eliminate heat loss to the surrounding gas molecules. The measurement method falls in the same category of comparative measurements as that based on the suspended Pt serpentine resistance thermometer
devices discussed in a previous section. According to the thermal circuit of Fig. 10, with a
fixed laser power ($Q_L$) absorbed by the cantilever and varying thermocouple temperature
($T_A$) at one end of the fiber,
\begin{equation}
R_{NF} = R_{cantilever} \frac{\Delta T_A - \Delta T_L}{\Delta T_L}
\end{equation}
where $R_{NF}$ and $R_{cantilever}$ are the thermal resistance of the nanofiber and the cantilever
sensor, and $\Delta T_A$ and $\Delta T_L$ are the external changes in thermocouple and cantilever tip
temperature ($T_L$) when $Q_L$ is kept constant. In principle, $R_{cantilever}$ can be obtained from
the change in the measured $T_L$ when $Q_L$ is varied, prior to the attachment of the nanofiber
to the cantilever. However, the temperature is nonuniform in the cantilever during the thermal measurement of the nanofiber, so that it is difficult to calibrate the cantilever defection
versus a uniform cantilever temperature and use the calibration result to determine $\Delta T_C$
during thermal measurement of the nanofiber. Instead, $Q_C$ is calibrated against the measured cantilever tip defection signal. Ideally, this calibration should be done in vacuum
before the nanofiber is attached to the cantilever, so that $Q_C$ equals the absorbed laser
power ($Q_L$) of the cantilever. Nevertheless, because $R_{NF}$ is three orders of magnitude
larger than $R_{cantilever}$, the heat loss through the nanofiber attached to the cantilever can
be neglected compared to $Q_C$, so that the calibration can be performed with the nanofiber
\begin{figure}[h]
 \centering
 \includegraphics[ width = .7\textwidth]{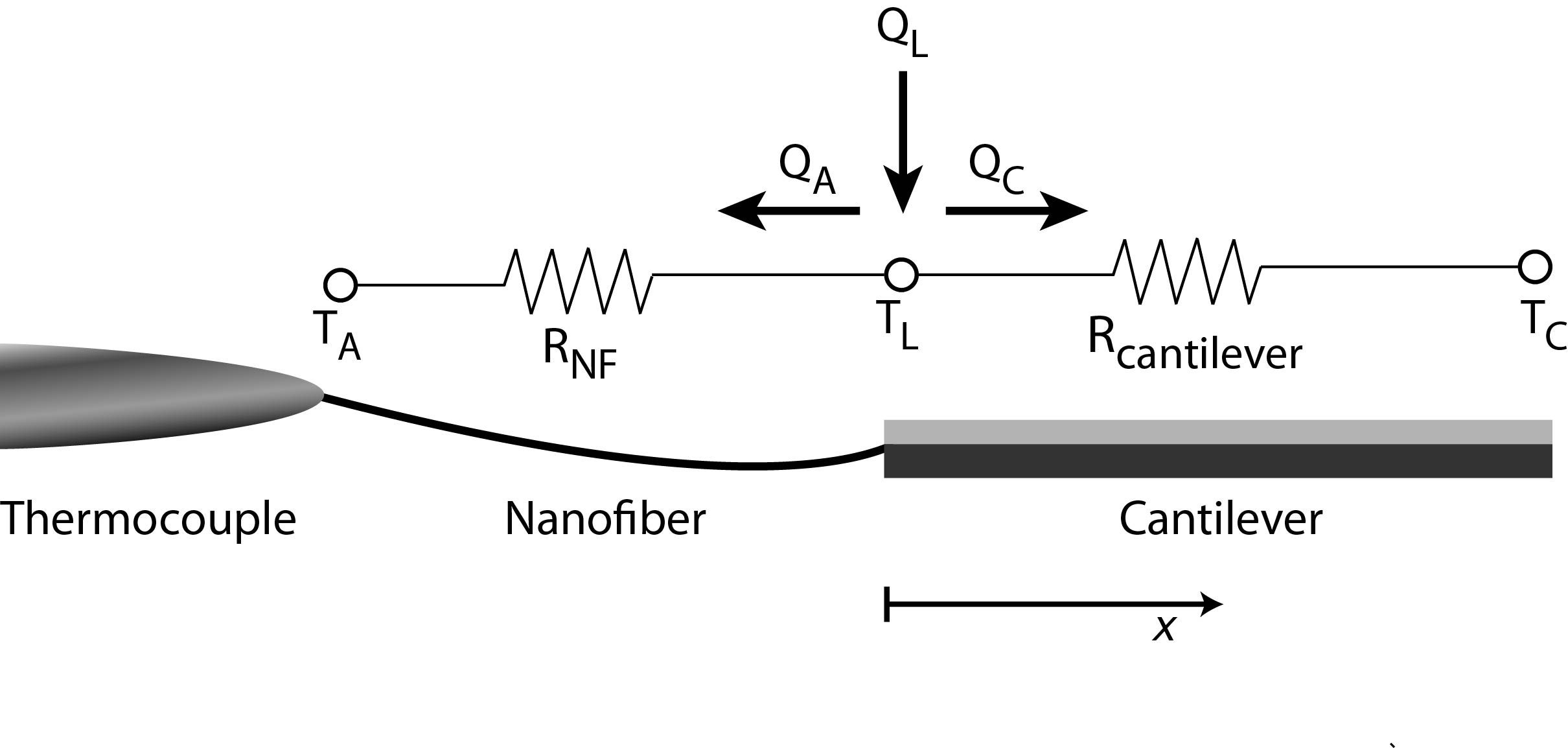}
 \caption{(a) Thermal circuit for the measurement of a polyethylene nanofiber using a bilayer cantilever and a heated thermocouple. $Q_L$ refers to the power absorbed by the incident
laser, and $T_L$ is the temperature of the cantilever at the laser spot, $R_{NF}$ and $R_{cantilever}$ are
the corresponding thermal resistances, and $T_A$ and $T_C$ are the temperatures of the thermocouple and cantilever base, respectively. Reproduced from Shen et al.$^{48}$ with permission
from Macmillan Publishers Ltd: Nature. Copyright 2010.
}
\end{figure}
attached to the cantilever. Their calibration yielded the ratio
\begin{equation}
\alpha_1 = \frac{\Delta Q_{L_{\Delta P}}}{\Delta B_{\Delta P}} \approx \frac{Q_{C_{\Delta P}}}{\Delta B_{\Delta P}}
\end{equation}
where $\Delta Q_{L_{\Delta P}}$, $\Delta Q_{C_{\Delta P}}$, and $\Delta B_{\Delta P}$ are the changes in $Q_L$, $Q_C$, and the cantilever defection signal, respectively, when the incident laser power ($P$) is varied and the thermocouple
temperature $T_A$ is held constant.

After calibration, $R_{NF}$ was determined by varying $T_A$ while maintaining constant
$Q_L$. The corresponding change in $Q_C$ can be obtained as $\Delta Q_{C_{\Delta T_A}} = \alpha_1 \Delta B_{\Delta T_A}$, where
$\Delta B_{\Delta T_A}$ is the corresponding change in the measured cantilever deflection signal. Hence,
\begin{equation}
R_{NF} = \frac{\Delta T_A - \Delta T_{L_{\Delta T_A}}}{\Delta Q_{C_{\Delta T_A}}} \approx \frac{\Delta T_A}{\Delta Q_{C_{\Delta T_A}}}
\end{equation}
where $\Delta T_{L_{\Delta T_A}}$ is the corresponding change in the cantilever tip temperature, and is about
three orders of magnitude smaller than $\Delta T_A$ because $R_{NF}$ is three orders of magnitude
larger than $R_{cantilever}$.

Because the $\sim$ 300 $\mu$m length of the nanofiber is three orders of magnitude larger than
its diameter of $\sim$ 130 nm, it is necessary to evaluate the radiation loss from its circumference in addition to heat conduction in the nanofiber. The nanofiber can be treated as a
radiative fin, with its thermal resistance related to the emissivity ($\epsilon$) and thermal conductivity ($\kappa$) as
\begin{equation}
R_{NF} = \frac{\sinh{\beta L}}{\sqrt{\pi^2 \epsilon \sigma \kappa T^3 D^3}}
\end{equation}
where $\beta = \sqrt{16\epsilon \sigma T^3 /D\kappa}$, $T$ is the average temperature of the system, and $\kappa$, $D$, and $L$
are the thermal conductivity, diameter, and length of the nanofiber, respectively. For $\beta \ll 1$,
Eq. (18) is reduced to the case of pure conduction, $R_{NF} = L/\kappa A$, which was used to obtain
a $\kappa$ value of $\sim$ 104 Wm$^{-1}$K$^{-1}$. For the corresponding $R_{NF}$ value and an emissivity $\epsilon = 0.1$,
Eq. (18) would yield a $\kappa$ value of 104.4  Wm$^{-1}$K$^{-1}$, just slightly higher than that determined
from neglecting radiation loss. The difference is negligible because of a small $\beta L$ value of 0.12.

A unique advantage of this measurement was that the bimaterial sensor was used for
both drawing the nanofiber and for measuring its thermal property. However, it is of interest to evaluate the temperature and heat flow rate measurement sensitivities of the bimaterial cantilever thermal sensor, and compare them with those achievable by resistance
thermometry. Based on the measurement data reported by Shen et al.,$^{48}$ the cantilever sensor is sensitive to about 3 K temperature change of $T_A$, which is equivalent to about $3 \times 10^{−3}$ K change in $T_L$. The measurement sensitivity in $Q_C$ is thus on the order of 30 nW/K
based on their reported $R_{cantilever}$ value on the order of $10^5$ K/W, and can potentially
be improved to about 0.3 nW/K if $R_{cantilever}$ is increased to the level of $10^7$ K/W found
for the thermal resistance of the supporting beams on suspended resistance thermometer
devices.$^{24}$ In addition, one source of error for this optical measurement is the uncertainty
in the absorbed laser power $Q_L$, which Shen et al.$^{48}$ determined to be as small as a $\sim$ 9\% of
the incident power and is found from subtracting the reflected and strayed beam intensity
from the incident intensity. The diffusely scattered laser power was thought to be small,
and neglected in their measurement. In comparison, it is feasible to determine the heating
in the cantilever rather conveniently and accurately if the heating is provided by electrical
heating of a doped Si or Pt-C resistor fabricated at the end of a cantilever.$^{49,50}$ This type of
resistance thermometer can achieve 3 mK temperature sensitivity with the use of lock-in
detection combined with a Wheatstone bridge. The resistance of the thermometer can also
be calibrated readily as a function of the cantilever temperature, and the method is free of
the complication caused by thermal drifting of the laser beam as well as cantilever deflection caused by mechanical strain applied by the nanofiber. This suggests that in addition
to bimaterial cantilevers, resistance thermometer cantilever sensors can also potentially
provide an attractive method for thermal measurements. However, the fabrication of the resistance thermometer cantilever sensors indeed requires additional steps of fabrication of
the doped Si or Pt-C resistance thermometer sensor at the end of the cantilever.

\section{Thermal Diffusivity Measurements with Doped Si Cantilever Resistance Thermometer Sensors}
Resistance thermometer cantilever sensors have been explored for thermal diffusivity measurements of polymer fibers by Demko et al.$^{49}$ The thermal sensor is a Si cantilever with
two heavily doped beams connected by a lightly doped Si region that acts as the resistance thermometer (see Fig. 11). The thermal measurement was conducted inside a SEM
chamber that was equipped with a nanomanipulator. During the measurement, a nanofiber
was suspended between the end of the Si cantilever and an aluminum support. After a
Joule heated micromanipulator probe was brought into contact with a point along the
suspended segment of the nanofiber, the time evolution of electrical resistance of the Si
thermometer was monitored by recording the voltage output from a Wheatstone bridge
circuit. If the thermal interface resistance at the two contacts to the nanofiber is ignored,
the fiber-probe contact point will maintain a constant temperature of the heated probe. Under this condition, a solution to the junction temperature between the nanofiber and the Si
cantilever can be found in Carslaw and Jaegers$^{51}$ for the case of two finite slabs in thermal
contact. The solution can be simplified for low-thermal conductivity and low-diffusivity
samples that satisfy the following conditions,
\begin{equation}
\frac{\kappa_S A_S}{\kappa_f A_f} \sqrt{\frac{\alpha_f}{\alpha_S}} = \frac{A_S}{A_f} \sqrt{\frac{\kappa_S \rho_S C_S}{\kappa_f \rho_f C_f}} \gg 1 \quad \textrm{and} \quad \sqrt{\frac{\alpha_f}{\alpha_S}} \ll \sqrt{\frac{\ell_f}{\ell_S}}
\end{equation}
where $\kappa$, $\alpha$, $\rho$, $C$, $A$, and $\ell$ refer to the thermal conductivity, diffusivity, density, heat capacity, cross section, and length of the sensor and fiber, denoted with the subscript $s$ and $f$,
respectively. In the simplified solution, the normalized temperature at the fiber-cantilever
junction located at $x = 0$ is given as
\begin{equation}
T^{*}(t) = \frac{\Delta T|_{x=0}}{\Delta T_m} \frac{\kappa_S A_S \ell_f}{\kappa_f A_f \ell_S} = 1 + 2 \sum\limits_{n=1}^{\infty} (-1)^{n} e^{-(\alpha_f n^2 \pi^2 t /\ell^2_{f})}
\end{equation}
where $\Delta T|_{x=0}$ and $\Delta T_m$ refer to the temperature rise at the fiber-cantilever interface and
that of the micromanipulator bought in contact at $x = −\ell_f$. The measured time evolution of
the sensor signal was fit to the functional form of Eq. (20) to extract the thermal diffusivity
of the sample.

For this thermal flash measurement of diffusivity to be sensitive and accurate, $\Delta T|_{x=0}$
needs to be large compared to the temperature sensitivity on the order $10^{-3}$ K of the resistance thermometer sensor. For $\Delta T_m$ of order 10 K, the ratio between the thermal resistance
of the fiber and that of the cantilever sensor, $R_f/R_s = \kappa_s A_s \ell_f /\kappa_f A_f \ell_s$, needs to be less than
the order of $10^4$. In addition, because the lightly doped resistance thermometer sensor has
a finite size, $l_{RT}$, the thermal time constant of the sensor needs to be considerably smaller than that of the fiber, that is, $l_{RT}^{2}/\alpha_s \ll \ell_f^2/\alpha_f$, in order to establish sufficient transient response of the sensor. In the work of Demko et al.,$^{49}$ the samples were glass fibers and
polyimide fibers with considerably lower thermal diffusivity than the Si sensor. For extending this method to a high-thermal diffusivity carbon nanotube of a finite length, the size
of the resistance thermometer may need to be reduced to be considerably smaller than the
length of the nanotube sample.

The above solutions are based on the absence of contact thermal resistance at the two
ends of the nanofiber. With the presence of a contact thermal resistance ($R_{f,c}$), the fiber
temperature at the fiber-cantilever contact ($T_{f,c}$) varies with time so that the above solution is not strictly applicable. However, when the thermal penetration depth increases
with time to be larger than $R_{f,c}A\kappa$, $T_{f,c}$ approaches a constant equal to the probe temperature. Hence, the above solution can still be used to fit the time variation of the sensor
signal after the initial response and extract the thermal diffusivity, although both the time
evolution and amplitude of the initial sensor response depend on the contact thermal resistance. Compared to a steady state measurement of the thermal conductivity, the error
caused by the contact thermal resistance is small but still present in the thermal flash measurement of the thermal diffusivity. However, accurate knowledge of the density and heat
capacity is necessary to convert the measured thermal diffusivity to thermal conductivity.

\begin{figure}[h]
 \centering
 \includegraphics[ width = .6\textwidth]{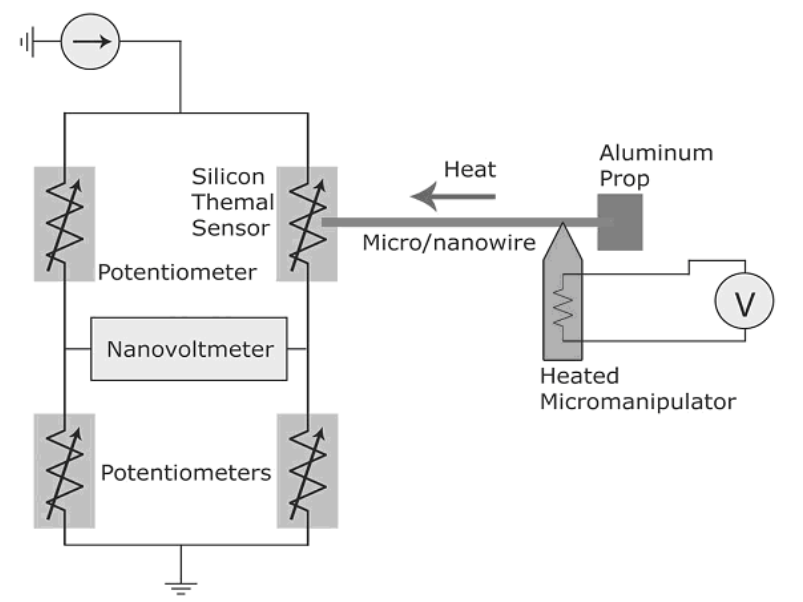}
 \caption{Schematic of the setup by Demko et al.$^{49}$ to measure the diffusivity of polyimide
nanofibers. Reproduced from Demko et al.$^{49}$ with permission from American Institute of
Physics. Copyright 2009.
}
\end{figure}
\section{Three Omega Techniques}
The 3$\omega$ technique for NW thermal conductivity measurement is a further development
of the celebrated method reported by Cahill$^{52}$ for thin film measurements. For NW measurements, the NW is suspended with four electrical contacts to the NW, as illustrated in
Fig. 12. A sinusoidal current at 1$\omega$ angular frequency is supplied to the NW sample, and
induces a temperature modulation at the 2$\omega$ frequency. The 2$\omega$ temperature fluctuation in
turn generates a 3$\omega$ component in the voltage drop measured across the suspended NW.
Measured with a lock-in amplifier, the rms amplitude of the 3$\omega$ voltage component is given
by
\begin{equation}
V_{3\omega} = \frac{4I^3 LR (dR/dT)}{\pi^4 \kappa A \sqrt{1 + (2\tau \omega)^2}}
\end{equation}
where the thermal time constant $\tau = L^2 /\pi^2 \alpha$, and $L$, $\alpha$, and $R$ are the length between the
voltage leads, thermal diffusivity, and electrical resistance of the NW sample. $I$ is the rms
amplitude of the time-dependent current, $I_0 \sin{\omega t}$. The experimental results can be fit to
Eq. (21) to extract the thermal conductivity and thermal time constant. The specific heat
can then be found from
\begin{equation}
C_p = \frac{\pi^2\tau \kappa}{\rho L^2}
\end{equation}

\begin{figure}[h]
 \centering
 \includegraphics[ width = .4\textwidth]{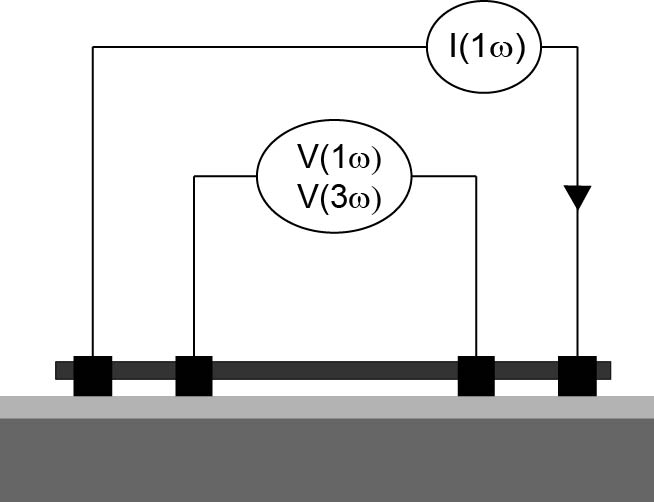}
 \caption{Schematic of the 3$\omega$ measurement of thermal conductivity and diffusivity of a
suspended nanowire.
}
\end{figure}
The thermal conductivity is often measured in the low frequency limit where the amplitude
in the temperature fluctuation is a maximum, and where the frequency dependent term in
Eq. (21) drops out. This is particularly important for the case of a long sample, where a
low frequency is needed so that the temperature modulation is larger than the measurement
sensitivity. Conversely, to obtain the specific heat, the measurement needs to be performed
at sufficiently high frequency compared to $1/\tau$,$^{53}$ so that $V_{3\omega}$ shows a clear frequency
dependence. For an individual carbon nanotube of a length of several micrometers and a
high thermal diffusivity, the frequency needs to be as high as 10 MHz in order to be able to
measure the specific heat. Measurements at such a high frequency can be prone to errors
caused by electrical capacitive coupling and other limitations in electronics.

The 3$\omega$ method has been used to measure the thermal conductivity and specific heat
of a number of conducting NWs and CNTs. Among the early 3$\omega$ measurements is the
work reported by Yi et al.$^{54}$ on the thermal conductivity and specific heat measurement of
1-2 mm long multiwalled carbon nanotube (MWCNT) bundles suspended over an etched
trench. The 3$\omega$ method has also been used for investigating thermal transport in Pt NWs at
cryogenic temperatures,$^{53}$ the Lorenz number in Ni NWs at low temperature,$^{55}$ the study
of phonon-surface scattering in metal-coated Si NWs,$^{56}$ the effect of twinned boundaries
on the thermal conductivity of InAs NWs below 6 K,$^{57}$ and individual defective CNTs.$^{58}$

A key advantage of using the self-heating 3$\omega$ method is that the transient measurement
gives not only the thermal conductivity, but also the specific heat and diffusivity. In addition, because the NW acts as a simple heater and sensor, the device fabrication is relatively
simple. Compared to steady state self-electrical heating and resistance thermometry methods, the 3$\omega$ method allows for sensitive frequency domain measurement of the induced
temperature rise in the NW sample. In addition to the 3$\omega$ method, the 1$\omega$ and 2$\omega$ frequency components of the voltage drop can be measured when there is a DC offset in the
applied modulated heating current. One example is the 1$\omega$ measurement of the DC temperature rise in the heating Pt serpentine of the resistance thermometer device of a large
thermal time constant.$^{24}$ A detailed analysis of the 1$\omega$, 2$\omega$, and 3$\omega$ detection methods is
given by Dames and Chen.$^{40}$

However, these self-electrical heating methods require that the NWs have a nearly constant temperature coefficient of resistance (TCR) in the measurement temperature range.
The presence of a temperature-dependent TCR requires a rather cumbersome analysis.
Moreover, one requirement of the validity of the self-heating methods is that the nonlinearity in the $I - V$ behavior should be due entirely to the change in lattice temperature,
and that the electrons are indeed in thermal equilibrium with the lattice. For example, in
the case of a short suspended CNT sample with scattering mean free path between optical
or acoustic phonons comparable to or longer than the suspended length, the local electron or optical phonon temperature can be considerably higher than the acoustic phonon
temperature. In this regime of highly nonequilibrium transport, it is not possible to convert the change in the measured electrical resistance to a calibrated temperature rise in the
sample.

\section{Transient Electrothermal Techniques}
Besides modulated and DC electrical self-heating, pulse electrical self-heating of the NW
sample can be employed for measuring the thermal conductivity and diffusivity of a suspended NW sample, with some potential advantages and disadvantages. Guo et al.$^{59}$ have
demonstrated this method with Pt wires, single-walled carbon nanotube (SWCNT) bundles, and Au-coated polyester fibers. In one of the implementations, the NW is suspended
across two copper electrodes, and a step function current is applied from time $t = 0$ to heat
the sample.$^{59}$ The electrical resistance of the wire was measured to determine the average
temperature rise in the suspended wire. With the radiation loss ignored, the normalized
average temperature rise of the wire is given as
\begin{equation}
T^{*}(t) = \frac{T(t) - T_0}{T(t\rightarrow \infty) - T_0} = \frac{96}{\pi^4} \sum\limits_{m=1}^\infty \frac{1 - \exp{\left ( -[(2m-1)^2\pi^2\alpha t / L^2)]\right )}}{(2m-1)^4}
\end{equation}
where the steady state temperature $T(t \rightarrow \infty) = T_0 + q_0 L^2 /12\kappa$, and $q_0$ is the heat
dissipation per unit volume in the nanowire, $L$ its length, and $\kappa$ is the thermal conductivity.
The thermal conductivity can be obtained from the steady state average temperature rise,
$\Delta T (t \rightarrow \infty)$.

At short times after the heating begins, when the heat transfer to the two ends of the
NW is sufficiently small, the temperature change in the wire depends linearly on time as,
$\Delta T = \Delta tq_0 /\rho C_p$, which agrees with the limiting case of Eq. (23), that is, $T^{*} = 12\Delta t\alpha/L^2$.
Hence, the diffusivity can be conveniently obtained from a linear fitting of the initial time
response of $T^{*}$. Alternatively, the entire $T^{*}$ versus $t$ response can be fitted with Eq. (23)
to obtain the thermal diffusivity. Besides a step increase in the heating current, similar
solutions can be found for a step decrease of the heating current.$^{60−62}$

In addition, it has been suggested by these works that the signal-to-noise ratio in pulsed
electrical heating is greater than that of the 3$\omega$ method, and has the advantage of a shorter
measurement time.$^{62}$ However, the pulsed heating/sensing methods described are suitable
only for materials with relatively low diffusivity and/or long lengths. For example, carbon
nanotubes, with diffusivity on the order of $2 \times 10^{−4}$ m$^2$/s and lengths on the order of
10 $\mu$m, have a thermal time constant, $L^2 /\alpha$, of 2 $\mu$s. In this case, the time scale of the
transient temperature response is on the same order as the rise time of most current sources,
i.e., $\sim$2 $\mu$s,$^{62}$ making it difficult to differentiate the contribution from the rise in electric
signal from the rise in sample temperature.

To address this issue, periodically modulated nanosecond laser pulses have been employed to heat the sample.$^{62}$ Under the assumption of a square pulse of uniform intensity
along the length of the suspended NW, the normalized average temperature rise in the
suspended NW was obtained as
\begin{equation}
T^{*} = \frac{T(t) - T_{min}}{T_{max}-T_{min}} = \frac{8}{\pi^2}\sum\limits_{m=1}^{\infty} \frac{\exp{\left (-[(2m-1)^2\pi^2 \alpha t / L^2]\right )}}{(2m-1)^2}
\end{equation}
where $T_{min}$ and $T_{max}$ are the minimum and maximum temperatures of the sample during
pulsed heating. The average temperature rise in the suspended NW was determined from
the electrical resistance of the sample measured with the use of a small DC current. By
fitting the measurement data with Eq. (24), this method has been employed to obtain the
thermal diffusivity of MWCNT bundles, Pt wires, and carbon fibers. However, without
knowledge of the optical absorption of the sample, the measurement does not yield the
thermal conductivity directly.

\section{Raman Thermometry-based Measurements}

Raman spectroscopy has been explored to probe thermal transport in CNTs,$^{63−65}$ graphene,$^{66−68}$ and GaAs NWs.$^{69}$ The sample temperature can be determined from the Raman spectrum in one of two ways. First, the Stokes to anti-Stokes intensity ratio can
provide the optical phonon temperature within the laser spot. However, the ratio depends
only on the zone center or zone boundary optical phonon populations, and the temperature
therefore corresponds to only the temperature of these Raman active modes. Moreover,the anti-Stokes peak is observable only after the sample temperature is heated to a high
temperature, and as high as 600 K for the case of graphene. On the other hand, as the temperature of the sample is increased, the Raman peaks become broadened and shifted due
to increased lattice anharmonicity. The shift in either the $2D$ band or $G$ band peak of the
Raman spectrum has been used to probe the temperature of graphene. However, the
temperature sensitivity is typically only on the order of 20-50 K.$^{65}$ In addition, charged
impurities and strain can contribute to the Raman peak shift. Therefore, it is important
that the strain and impurity scattering remain essentially constant throughout the measured
temperature range, to ensure the peak shift can be attributable to temperature change alone.
A further complication in Raman-based thermometry is the possible presence of nonequilibrium phonon transport among the acoustic modes and the Raman-excited optical modes.
If acoustic modes, with very long mean free path, do not participate in the relaxation of the
absorbed photons and hot electrons in the optically heated or electrically biased sample, or
do not interact effectively with the optically excited Raman modes, they will be at a lower
temperature than the optical phonon or electron populations. This is particularly important
for very short CNTs when the optical phonon transport is quasi-ballistic.

Despite these experimental complications, Raman measurements on graphene, CNTs,
and semiconducting NWs have yielded useful insight into phonon transport in nanostructures. Hsu et al.$^{65}$ have studied the heat transfer in suspended SWCNTs by recording the
shift in the $G$ peak as the laser spot was scanned along the nanotube suspended across a
trench. Based on a separate calibration, the measured $G$ peak shift was converted to the
local temperature rise in the CNT. The obtained temperature profile suggested diffusive
phonon transport in the defective CNT sample can be fit to the parabolic temperature profile with a curvature given by $Q/\kappa AL$, where $Q$ is the optical heating rate absorbed by
the nanotube, $\kappa$, $A$, and $L$ are the thermal conductivity, cross section, and length of the nanotube. Because $Q$ is unknown, this measurement can provide only the ratio of contact
resistance to sample resistance of the suspended CNT. The ratio ranged from 0.02 to as
high as 17. In another work on GaAs NWs, Soini et al.$^{69}$ used an ab initio finite difference
simulation to extract the $Q$ term, allowing for the calculation of the thermal conductivity from the
temperature profile of the NW.

In order to experimentally determine the optical absorption in a CNT, Hsu et al.$^{70}$
focused a Raman laser beam onto a $\sim$ 400 nm segment of a $\sim$ 10 $\mu$m-long SWCNT bundle
suspended between two suspended microscale Pt resistance thermometers, and measured
the temperature rises ($\Delta T_1$ and $\Delta T_2$) at two thermometers. The laser power absorbed by the nanotube was taken to be the $Q = (\Delta T_1 +\Delta T_2 )/R_b$, where $R_b$ is the thermal resistance
of the supporting beams for each thermometer. With the contact thermal resistance ignored,
the temperature of the CNT at the laser spot was determined form the red shift of the $G$
band, and used together with $Q$ to obtain the thermal conductivity.
\begin{figure}[h]
 \centering
 \includegraphics[ width = .7\textwidth]{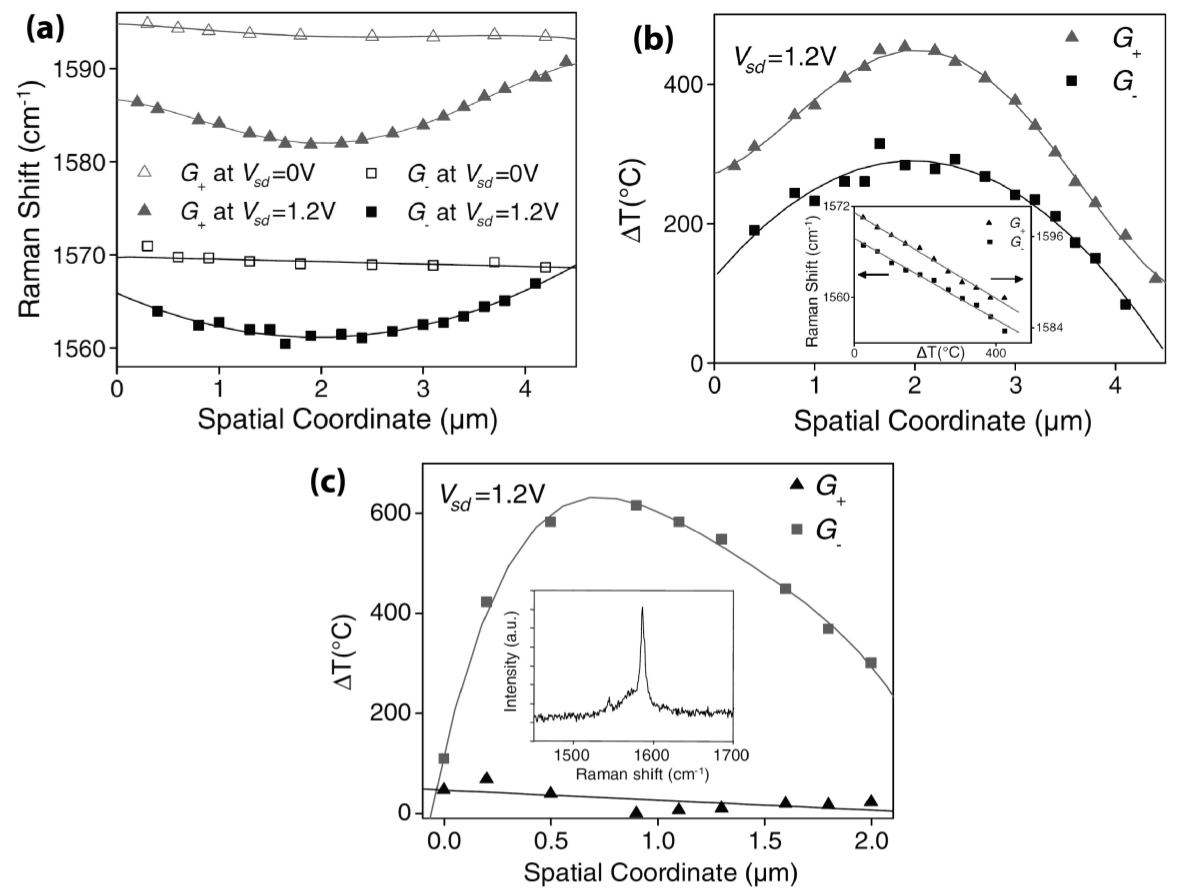}
 \caption{(a) Raman shift of a 5 $\mu$m-length suspended CNT and (b) corresponding temperature profile for the $G^{−}$ and $G^+$ phonon modes. (c) Temperature profile of a 2 $\mu$m-length
suspended CNT. The $G^+$ mode shows no downshift with applied electric field. Reproduced from Bushmaker et al.$^{72}$ with permission from American Chemical Society. Copyright 2007.
}
\end{figure}

Compared to optical heating, the electrical heating rate can be obtained readily. However, it is nontrivial to extract a thermal property from electrically biased, high-quality
CNTs because preferential coupling between the hot electrons with optical phonons can
drive the phonon populations out of local thermal equilibrium.$^{71−73}$ For example, Fig. 13
(Ref. 71) shows the Raman shift and the temperature profile for a 5 $\mu$m-long suspended
CNT [Figs. 13(a) and 13(b)] and another 2 $\mu$m-long [Fig. 13(c)] suspended CNT, measured
with 0 and 1.2 V bias voltage. The $G^+$ and $G^-$ phonon peaks were attributed to the transverse optical (TO) and longitudinal optical (LO) phonon modes of metallic nanotubes. For
the long nanotube, electrical heated with 1.2 V, both phonon modes are excited to a similar temperature by the energetic electrons. However, for the short nanotube in Fig. 13(c),
only one of the peaks downshifts, in this case the $G^-$ peak, with the applied electric bias.
The lower temperature profile in Fig. 13(c) is expected to be an upper bound to the lattice
temperature, and thus while the $G^-$ mode exists at a high effective temperature, the lattice remains at room temperature. Such local nonequilibrium in the phonon temperature in
short CNTs is explained by the slow decay of the hot optical phonons into other phonon
modes.

The nonequilibrium issue is less a problem in longer, suspended CNTs. The thermal
conductivities of electrically biased, long suspended SWCNTs and MWCNTs have been
obtained by Li et al.$^{63}$ with the use of micro-Raman spectroscopy. The CNTs were grown
over 40 $\mu$m-deep trenches with two patterned Mo electrodes on either end of the CNT,
allowing for electrical heating of the CNT. Based on the Raman $G$ peak position shift, the
CNT temperature was probed with the Raman laser at the ends and middle of the suspended
sample, and the difference in temperature was used to find the thermal conductivity from
the measured electrical heating power and the dimensions of the sample. The effects of
contact resistance can be eliminated from the temperature rise measured by the contact.
One concern in the measurement of the long suspended nanotube of a small diameter is the
radiation loss. Li et al.$^{63}$ estimated that the radiation loss was $<$ 1\% of the heating power.
The small radiation loss is due to the small emissivity and large thermal conductivity of
the high-quality CNT. Despite these progresses, the limited temperature sensitivity of the
Raman techniques requires that the CNT be heated to a rather high temperature during
the measurement, and the temperature dependence of the thermal conductivity cannot be
resolved with sufficient accuracy, especially at low temperatures.

Further work by Hsu et al.$^{74}$ has investigated the dependence of the gas environment
on the heat dissipation from electrically biased suspended SWCNTs. They found a much
lower temperature rise in CNTs heated within a gas environment, with 50 - 60\% of the
Joule heating being conducted away by the surrounding gas. This effect was particularly
pronounced for a CO$_2$ environment and other polyatomic gas molecules with lower thermal
conductivity. The results are attributed to the coupling between the hot surface optical
phonons of the CNT and the molecular vibration modes of the surrounding polyatomic
molecules. Using a two-laser technique, Hsu et al.$^{75}$ further measured the heat transfer
coefficient between SWCNT bundles and the surrounding air environment. A suspended
CNT bundle with diameters on the order of tens of microns and suspended lengths as long
as 89 $\mu$m was heated locally with a 750 nm spot size laser. A second laser was incident on
the CNT for probing the spatially resolved Raman spectra with resolution up to 1 $\mu$m. With
the same laser heating power, the maximum temperature increase was 150 K for irradiation
in air, compared to 275 K in vacuum. The heat transfer coefficient between the CNT and
surroundings was found from the fin heat transfer solution, and ranged from 0.15 to $7.91 \times
10^4$ W/m$^2$K. The heat dissipation to the surrounding air is found to be the dominant path
of heat transfer for CNT bundles longer than 7 $\mu$m. The results suggest the importance
of performing thermal conductivity measurements of CNTs, especially defective CNTs of
low thermal conductivity, in a vacuum environment.

\section{Time Domain Thermoreflectance}
Time domain thermoreflectance (TDTR) has been an increasingly popular method for characterization of the thermal property of thin films and interfaces. There have also been reported TDTR measurements on aligned InAs NW-polymer composites,$^{76}$ from which the
average thermal conductivity of the individual NWs in the composite can be obtained. InAs
NW arrays were grown by chemical beam epitaxy on an InAs substrate, followed by embedding the entire NW array in polymethyl methacrylate (PMMA). To ensure contact with
the deposited 90 nm Al film, the PMMA was etched by ozone plasma until the tips of the
vertically aligned NWs were just above the PMMA surface. The estimated void fraction
in the NW composite was estimated to be $<$ 2\%. A laser pulse with duration $<$ 0.3 ps was
used to heat the sample, while a probe pulse was used to measure the change in reflectance
of the top Al film. The thermal penetration depth of the heating pulse depends on its frequency $f$ as $d = \sqrt{\kappa_C/\pi C_C f}$, where $\kappa_C$ is the composite thermal conductivity, and $C_C$
is the composite specific heat. It was found that the thermal conductivity of the composite
depended on this modulation frequency, which they attribute to a transition between two
limits of the NW/PMMA and NW/Al interface conductance: an effective medium limit at
low frequency and a two-temperature limit in which the NWs are thermally insulated from
the PMMA at high frequencies. The thermal conductivity of the InAs component of the
sample can be found from the low-frequency thermal conductivity of the composite from
the effective medium theory,
\begin{equation}
\kappa_C = x\kappa_{NW} + (1-x)\kappa_{PMMA}
\end{equation}
with the thermal conductivity of PMMA ($\kappa_{P M M A}$) determined from a separate experiment, and the packing fraction of NWs, $x$, determined from SEM. This results in a room
temperature thermal conductivity for individual InAs NWs of 5.3  Wm$^{-1}$K$^{-1}$, which is consistent
with another measurements of individual InAs NWs with the use of a suspended resistance
thermometer device.$^{77}$

\section{Summary and Outlook}
A number of experimental methods have been explored in the past decade to address the
challenge in thermal and thermoelectric transport measurements of NWs and nanotubes.
As discussed above, these methods are based on measurements of either steady state or
transient temperature rise and heat flow in NW samples that are heated either externally
or internally. For temperature measurements in these experiments, one popular technique
is based on microscale resistance thermometers, including suspended Pt serpentines, metal
hotwires, or the use of the NW itself as a resistance thermometer. Bimaterial cantilevers
have also been employed as temperature sensors in thermal conductivity measurement of
NWs. In addition, noncontact optical measurement techniques such as micro-Raman spectroscopy and thermal reflectance measurements have been investigated for determining
the temperature drop or temperature distribution in the NW sample. Although thermocouple sensors are commonly used in bulk thermal conductivity measurements and there
have been reports of encouraging progress in quantitative nanoscale temperature mapping
with the use of scanning thermal microscopy (SThM) techniques based on microfabricated thermocouple probes,$^{78-80}$ thermocouple sensors have remained to be explored for
NW thermal property measurements. Among these temperature measurement techniques,
resistance thermometers and bimaterial cantilevers can provide superior temperature sensitivity but limited spatial resolution compared to measurements based on microfabricated
thermocouples and optical thermometry techniques. The development of new temperature
measurement techniques with enhanced temperature sensitivity and spatial resolution and
reduced parasitic heat loss are essential for addressing several challenging problems still
present in existing techniques for NW measurements.

One common problem in existing techniques is the error caused by contact thermal resistance. Efforts have been made to overcome this problem by determining the respective
temperature drop across the suspended segment of the NW and at the contacts to the NW
with the use of the four-probe thermoelectric measurement technique, measuring
the length dependence of the NW thermal resistance, or measuring the spatial profile of the temperature distribution along the NW with the use of the micro-Raman spectroscopy technique.
Quantitative nanoscale SThM and other novel thermal imaging techniques that provide
improved spatial resolution and temperature sensitivity compared to micro-Raman spectroscopy may find use in addressing the contact thermal resistance problem. Furthermore,
in transient measurements of thermal diffusivity, the measured thermal time constant of
the sample is relatively insensitive to the contact thermal resistance,$^{49}$ providing another
powerful set of techniques for thermal characterization.

Another challenging problem in thermal measurement of individual NWs is the determination of the actual heat flow rate in the sample. Thermal measurement techniques based
on optical heating of the sample can benefit from new methods for accurate determination
of the optical absorption of the sample. In addition, radiation loss from long NWs of small
diameter needs to be evaluated adequately.

Besides the challenges in measuring the contact thermal resistance and the actual heat
flow rate in the sample, accurate determinations of the cross-sectional area and crystal
structure of the NW sample are essential for obtaining high-fidelity thermal conductivity
values and for correlating the results with the crystal structure. For a NW sample with an
irregular or rough surface, the cross section and thermal conductivity are not well defined.
Hence, comparing the thermal conductivity between rough and smooth NWs is nontrivial.

Although progress has been made in obtaining thermal conductivity, Seebeck coefficient, and electrical conductivity of individual NWs, there is still a lack of experimental
methods for probing the fundamental transport parameters in individual NWs. There have
been some efforts in extracting the electrochemical potential, charge carrier concentration,
and mobility from field-effect measurements$^{15,81,82}$ or thermoelectric measurements$^{27,83}$ of individual NWs. However, measurements of phonon dispersion and lifetime in NWs
have remained to be explored, and have not been achieved with the use of inelastic neutron
scattering and X-ray scattering techniques established for bulk crystals, because the crystal
size needs to be sufficiently large for these techniques. Knowledge of these fundamental
phonon and electron transport parameters are critically needed to understand the many intriguing experimental results of thermal conductivity, Seebeck coefficient, and electrical
conductivity of individual NWs. Hence, the current limited experimental capability calls
for innovative approaches for probing thermal and thermoelectric transport in individual
NWs.

\section{Acknowledgement}
The authors thank Profs. Alexis Abramson, Chris Dames, and Deyu Li, and Drs. Michael
T. Pettes and Arden L. Moore for helpful discussion. The authors also acknowledge funding support from the National Science Foundation (NSF), Department of Energy, and Office of Naval Research. AW is supported by a NSF Graduate Research Fellowship.

\section{References}
\begin{enumerate}
\item Y. Li, F. Qian, J. Xiang, and C. M. Lieber, Nanowire electronic and optoelectronic devices, \emph{Mater. Today}, \textbf{9}:18-27, 2006.
\item R. Agarwal and C. M. Lieber, Semiconductor nanowires: optics and optoelectronics, \emph{Appl.
Phys. A}, \textbf{85}:209-215, 2006.
\item R. Yan, D. Gargas, and P. Yang, Nanowire photonics, \emph{Nat. Photon.}, \textbf{3}:569-576, 2009.
\item R. Martel, T. Schmidt, H. R. Shea, T. Hertel, and P. Avouris, Single- and multi-wall carbon
nanotube field-effect transistors, \emph{Appl. Phys. Lett.}, \textbf{73}:2447-2449, 1998.
\item S. J. Tans, A. R. M. Verschueren, and C. Dekker, Room-temperature transistor based on a single
carbon nanotube, \emph{Nature}, \textbf{393}:49-52, 1998.
\item T. Someya, J. Small, P. Kim, C. Nuckolls, and J. T. Yardley, Alcohol vapor sensors based on
single-walled carbon nanotube field effect transistors, \emph{Nano Lett.}, \textbf{3}:877-881, 2003.
\item Y. Cui, Z. Zhong, D. Wang, W. U. Wang, and C. M. Lieber, High performance silicon nanowire
field effect transistors, \emph{Nano Lett.}, \textbf{3}:149-152, 2003.
\item D. Hisamoto, W.-C. Lee, J. Kedzierski, H. Takeuchi, K. Asano, C. Kuo, E. Anderson,
T.-J. King, J. Bokor, and C. Hu, FinFET-a self-aligned double-gate MOSFET scalable to 20
nm, \emph{IEEE Trans. Electron Devices}, \textbf{47}:2320-2325, 2000.
\item X. Huang, W.-C. Lee, C. Kuo, D. Hisamoto, L. Chang, J. Kedzierski, E. Anderson, H. Takeuchi,
Y.-K. Choi, K. Asano, V. Subramanian, T.-J. King, J. Bokor, and C. Hu, Sub-50 nm P-channel
FinFET, \emph{IEEE Trans. Electron Devices}, \textbf{48}:880-886, 2001.
\item E. Pop and K. E. Goodson, Thermal phenomena in nanoscale transistors, \emph{J. Electron. Packag.},
\textbf{128}:102-108, 2006.
\item D. G. Cahill, W. K. Ford, K. E. Goodson, G. D. Mahan, A. Majumdar, H. J. Maris, R. Merlin,
and S. R. Phillpot, Nanoscale thermal transport, \emph{J. Appl. Phys.}, \textbf{93}:793-818, 2003.
\item C. J. Vineis, A. Shakouri, A. Majumdar, and M. G. Kanatzidis, Nanostructured thermoelectrics:
Big efficiency gains from small features, \emph{Adv. Mater.}, \textbf{22}:3970-3980, 2010.
\item L. D. Hicks and M. S. Dresselhaus, Effect of quantum-well structures on the thermoelectric
figure of merit, \emph{Phys. Rev. B}, \textbf{47}:12727-12731, 1993.
\item G. D. Mahan and J. O. Sofo, The best thermoelectric, Proc. Natl. Acad. Sci., 93:7436-7439,
1996.
\item J. Heremans and C. M. Thrush, Thermoelectric power of bismuth nanowires, Phys. Rev. B,
59:12579-12583, 1999.
\item A. Henry, G. Chen, S. J. Plimpton, and A. Thompson, 1D-to-3D transition of phonon heat
conduction in polyethylene using molecular dynamics simulations, \emph{Phys. Rev. B}, \textbf{82}:144308,
2010.
\item R. P. Tye, \emph{Thermal Conductivity}, Academic Press, New York, 1969.
\item T. S. Tighe, J. M. Worlock, and M. L. Roukes, Direct thermal conductance measurements on
suspended monocrystalline nanostructures, \emph{Appl. Phys. Lett.}, \textbf{70}:2687-2689, 1997.
\item K. Schwab, E. A. Henriksen, J. M. Worlock, and M. L. Roukes, Measurement of the quantum
of thermal conductance, \emph{Nature}, \textbf{404}:974-977, 2000.
\item K. Schwab, J. L. Arlett, J. M. Worlock, and M. L. Roukes, Thermal conductance through
discrete quantum channels, \emph{Physica E}, \textbf{9}:60-68, 2001.
\item L. G. C. Rego and G. Kirczenow, Quantized thermal conductance of dielectric quantum wires,
\emph{Phys. Rev. Lett.}, \textbf{81}:232-235, 1998.
\item L. Shi, Mesoscopic Thermophysical measurements of microstructures and carbon nanotubes,
PhD Thesis, University of California-Berkeley, 2001.
\item P. Kim, L. Shi, A. Majumdar, and P. L. McEuen, Thermal transport measurements of individual
multiwalled nanotubes, \emph{Phys. Rev. Lett.}, \textbf{87}:215502, 2001.
\item L. Shi, D. Li, C. Yu, W. Jang, D. Kim, Z. Yao, P. Kim, and A. Majumdar, Measuring thermal and
thermoelectric properties of one-dimensional nanostructures using a microfabricated device, \emph{J.
Heat Transfer}, \textbf{125}:881-888, 2003.
\item D. Li, Y. Wu, P. Kim, L. Shi, P. Yang, and A. Majumdar, Thermal conductivity of individual
silicon nanowires, \emph{Appl. Phys. Lett.}, \textbf{83}:2934-2936, 2003.
\item A. Mavrokefalos, M. T. Pettes, F. Zhou, and L. Shi, Four-probe measurements of the in-plane
thermoelectric properties of nanofilms, \emph{Rev. Sci. Instrum.}, \textbf{78}:034901, 2007.
\item F. Zhou, J. Szczech, M. T. Pettes, A. L. Moore, S. Jin, and L. Shi, Determination of transport properties in chromium disilicide nanowires via combined thermoelectric and structural
characterizations, \emph{Nano Lett.}, \textbf{7}:1649-1654, 2007.
\item J. Yang, Y. Yang, S. W. Waltermire, X. Wu, H. Zhang, T. Gutu, Y. Jiang, Y. Chen, A. A. Zinn,
R. Prasher, T. T. Xu, and D. Li, Enhanced and switchable nanoscale thermal conduction due to
van der Waals interfaces, \emph{Nat. Nano}, \textbf{7}:91-95, 2012.
\item A. Mavrokefalos, A. L. Moore, M. T. Pettes, L. Shi, W. Wang, and X. Li, Thermoelectric and
structural characterizations of individual electrodeposited bismuth telluride nanowires, \emph{J. Appl.
Phys.}, \textbf{105}:104318, 2009.
\item M. T. Pettes and L. Shi, Thermal and structural characterizations of individual single-, double-,
and multi-walled carbon nanotubes, \emph{Adv. Funct. Mater.}, \textbf{19}:3918-3925, 2009.
\item J. Zhou, C. Jin, J. H. Seol, X. Li, and L. Shi, Thermoelectric properties of individual electrodeposited bismuth telluride nanowires, \emph{Appl. Phys. Lett.}, \textbf{87}:133109, 2005.
\item D. Tham, C. Y. Nam, and J. E. Fischer, Microstructure and composition of focused-ion-beam-deposited Pt contacts to GaN nanowires, \emph{Adv. Mater.}, \textbf{18}:290-294, 2006.
\item Y. Long, Z. Chen, W. Wang, F. Bai, A. Jin, and C. Gu, Electrical conductivity of single CdS
nanowire synthesized by aqueous chemical growth, \emph{Appl. Phys. Lett.}, \textbf{86}:153102, 2005.
\item A. Motayed, A. V. Davydov, M. D. Vaudin, I. Levin, J. Melngailis, and S. N. Mohammad,
Fabrication of GaN-based nanoscale device structures utilizing focused ion beam induced Pt
deposition, \emph{J. Appl. Phys.}, \textbf{100}:024306, 2006.
\item V. Gopal, E. A. Stach, V. R. Radmilovic, and I. A. Mowat, Metal delocalization and surface
decoration in direct-write nanolithography by electron beam induced deposition, \emph{Appl. Phys.
Lett.}, \textbf{85}:49-51, 2004.
\item J. Tang, H.-T. Wang, D. H. Lee, M. Fardy, Z. Huo, T. P. Russell, and P. Yang, Holey silicon as
an efficient thermoelectric material, \emph{Nano Lett.}, \textbf{10}:4279-4283, 2010.
\item A. Weathers, A. L. Moore, M. T. Pettes, D. Salta, J. Kim, K. Dick, L. Samuelson, H. Linke,
P. Caroff, and L. Shi, Phonon transport and thermoelectricity in defect-engineered InAs
nanowires, \emph{Proc. MRS Spring Meeting}, vol. 1404, MRS Online Library, Cambridge, 2012.
\item C. Yu, S. Saha, J. Zhou, L. Shi, A. M. Cassell, B. A. Cruden, Q. Ngo, and J. Li, Thermal
contact resistance and thermal conductivity of a carbon nanofiber, \emph{J. Heat Transfer}, \textbf{128}:234-
239, 2006.
\item A. L. Moore and L. Shi, On errors in thermal conductivity measurements of suspended and
supported nanowires using micro-thermometer devices from low to high temperatures, \emph{Meas.
Sci. Technol.}, \textbf{22}:015103, 2011.
\item C. Dames and G. Chen, 1 omega, 2 omega, and 3 omega methods for measurements of thermal
properties, \emph{Rev. Sci. Instrum.}, \textbf{76}:124902, 2005.
\item M. C. Wingert, Z. C. Y. Chen, E. Dechaumphai, J. Moon, J.-H. Kim, J. Xiang, and R. Chen,
Thermal conductivity of Ge and Ge-Si core-shell nanowires in the phonon confinement
regime, \emph{Nano Lett.}, \textbf{11}:5507-5513, 2011.
\item N. Ashcroft and N. D. Mermin, \emph{Solid State Physics}, Brooks/Cole, Belmont, CA, pp. 523-528,
1976.
\item M. P. Marder, \emph{Condensed Matter Phys.}, Wiley, Hoboken, 2000.
\item K. Hippalgaonkar, B. Huang, R. Chen, K. Sawyer, P. Ercius, and A. Majumdar, Fabrication of
microdevices with integrated nanowires for investigating low-dimensional phonon transport,
\emph{Nano Lett.}, \textbf{10}:4341-4348, 2010.
\item J. Yang, Y. Yang, S. W. Waltermire, T. Gutu, A. A. Zinn, T. T. Xu, Y. Chen, and D. Li, Measurement of the intrinsic thermal conductivity of a multiwalled carbon nanotube and its contact
thermal resistance with the substrate, \emph{Small}, \textbf{7}:2334-2340, 2011.
\item M. Fujii, X. Zhang, H. Xie, H. Ago, K. Takahashi, T. Ikuta, H. Abe, and T. Shimizu, Measuring
the thermal conductivity of a single carbon nanotube, \emph{Phys. Rev. Lett.}, \textbf{95}:065502, 2005.
\item C. Dames, S. Chen, C. T. Harris, J. Y. Huang, Z. F. Ren, M. S. Dresselhaus, and G. Chen, A
hot-wire probe for thermal measurements of nanowires and nanotubes inside a transmission
electron microscope, \emph{Rev. Sci. Instrum.}, \textbf{78}:104903, 2007.
\item S. Shen, A. Henry, J. Tong, R. Zheng, and G. Chen, Polyethylene nanofibres with very high
thermal conductivities, \emph{Nat. Nanotechnol.}, \textbf{5}:251-255, 2010.
\item M. T. Demko, Z. Dai, H. Yan, W. P. King, M. Cakmak, and A. R. Abramson, Application
of thethermal fash technique for low thermal diffusivity micro/nanofibers, \emph{Rev. Sci. Instrum.},
\textbf{80}:036103, 2009.
\item A. Narayanaswamy, S. Shen, and G. Chen, Near-field radiative heat transfer between a sphere
and a substrate, \emph{Phys. Rev. B}, \textbf{78}:115303, 2008.
\item H. S. Carslaw and J. C. Jaeger, \emph{Heat Conduction in Solids}, 2nd ed., Oxford University Press,
New York, 1959.
\item D. G. Cahill, Thermal conductivity measurement from 30 to 750 K: the 3 omega method, \emph{Rev.
Sci. Instrum.}, \textbf{61}:802-808, 1990.
\item L. Lu, W. Yi, and D. L. Zhang, 3 omega method for specific heat and thermal conductivity
measurements, \emph{Rev. Sci. Instrum.}, \textbf{72}:2996-3003, 2001.
\item W. Yi, L. Lu, Z. Dian-lin, Z. W. Pan, and S. S. Xie, Linear specific heat of carbon nanotubes,
\emph{Phys. Rev. B}, \textbf{59}:R9015-R9018, 1999.
\item M. N. Ou, T. J. Yang, S. R. Harutyunyan, Y. Y. Chen, C. D. Chen, and S. J. Lai, Electrical and
thermal transport in single nickel nanowire, \emph{Appl. Phys. Lett.}, \textbf{92}:063101, 2008.
\item J. S. Heron, T. Fournier, N. Mingo, and O. Bourgeois, Mesoscopic size effects on the thermal
conductance of silicon nanowire, \emph{Nano Lett.}, \textbf{9}:1861-1865, 2009.
\item S. Dhara, H. S. Solanki, A. P. R, V. Singh, S. Sengupta, B. A. Chalke, A. Dhar, M. Gokhale,
A. Bhattacharya, and M. M. Deshmukh, Tunable thermal conductivity in defect engineered
nanowires at low temperatures, \emph{Phys. Rev. B}, \textbf{84}:121307, 2011.
\item T.-Y. Choi, D. Poulikakos, J. Tharian, and U. Sennhauser, Measurement of the thermal conductivity of individual carbon nanotubes by the four-point three-$\omega$ method, \emph{Nano Lett.}, \textbf{6}:1589-
1593, 2006.
\item J. Guo, X. Wang, and T. Wang, Thermal characterization of microscale conductive and non-conductive wires using transient electrothermal technique, \emph{J. Appl. Phys.}, \textbf{101}:063537, 2007.
\item B. Feng, W. Ma, Z. Li, and X. Zhang, Simultaneous measurements of the specific heat and
thermal conductivity of suspended thin samples by transient electrothermal method, \emph{Rev. Sci.
Instrum.}, \textbf{80}:064901, 2009.
\item X. Huang, J. Wang, G. Eres, and X. Wang, Thermophysical properties of multi-wall carbon
nanotube bundles at elevated temperatures up to 830 K, \emph{Carbon}, \textbf{49}:1680-1691, 2011.
\item J. Guo, X. Wang, D. B. Geohegan, G. Eres, and C. Vincent, Development of pulsed laser-
assisted thermal relaxation technique for thermal characterization of microscale wires, \emph{J. Appl.
Phys.}, \textbf{103}:113505, 2008.
\item Q. Li, C. Liu, X. Wang, and S. Fan, Measuring the thermal conductivity of individual carbon
nanotubes by the Raman shift method, \emph{Nanotechnology}, \textbf{20}:145702, 2009.
\item Y. Yue, G. Eres, X. Wang, and L. Guo, Characterization of thermal transport in micro/nanoscale
wires by steady-state electro-Raman-thermal technique, \emph{Appl. Phys. A}, \textbf{97}:19-23, 2009.
\item I. K. Hsu, R. Kumar, A. Bushmaker, S. B. Cronin, M. T. Pettes, L. Shi, T. Brintlinger,
M. S. Fuhrer, and J. Cumings, Optical measurement of thermal transport in suspended carbon
nanotubes, \emph{Appl. Phys. Lett.}, \textbf{92}:063119, 2008.
\item A. A. Balandin, S. Ghosh, W. Z. Bao, I. Calizo, D. Teweldebrhan, F. Miao, and C. N. Lau,
Superior thermal conductivity of single-layer graphene, 
\emph{Nano Lett.}, \textbf{8}:902-907, 2008.
\item W. W. Cai, A. L. Moore, Y. W. Zhu, X. S. Li, S. S. Chen, L. Shi, and R. S. Ruoff, Thermal
transport in suspended and supported monolayer graphene grown by chemical vapor deposition, \emph{Nano Lett.}, \textbf{10}:1645-1651, 2010.
\item C. Faugeras, B. Faugeras, M. Orlita, M. Potemski, R. R. Nair, and A. K. Geim, Thermal conductivity of graphene in corbino membrane geometry, \emph{ACS Nano}, \textbf{4}:1889-1892, 2010.
\item M. Soini, I. Zardo, E. Uccelli, S. Funk, G. Koblmuller, A. Fontcuberta i Morral, and G. Abstreiter, Thermal conductivity of GaAs nanowires studied by micro-Raman spectroscopy combined
with laser heating, \emph{Appl. Phys. Lett.}, \textbf{97}:263107, 2010.
\item I. K. Hsu, M. T. Pettes, A. Bushmaker, M. Aykol, L. Shi, and S. B. Cronin, Optical absorption
and thermal transport of individual suspended carbon nanotube bundles, \emph{Nano Lett.}, \textbf{9}:590-
594, 2009.
\item V. V. Deshpande, S. Hsieh, A. W. Bushmaker, M. Bockrath, and S. B. Cronin, Spatially re-
solved temperature measurements of electrically heated carbon nanotubes, 
\emph{Phys. Rev. Lett.},
\textbf{102}:105501, 2009.
\item A. W. Bushmaker, V. V. Deshpande, M. W. Bockrath, and S. B. Cronin, Direct observation of
mode selective electron-phonon coupling in suspended carbon nanotubes, \emph{Nano Lett.}, \textbf{7}:3618-
3622, 2007.
\item M. Oron-Carl and R. Krupke, Raman spectroscopic evidence for hot-phonon generation in
electrically biased carbon nanotubes, \emph{Phys. Rev. Lett.}, \textbf{100}:127401, 2008.
\item I. K. Hsu, M. T. Pettes, M. Aykol, L. Shi, and S. B. Cronin, The effect of gas environment onelectrical heating in suspended carbon nanotubes, \emph{J. Appl. Phys.}, \textbf{108}:084307, 2010.
\item I. K. Hsu, M. T. Pettes, M. Aykol, C.-C. Chang, W.-H. Hung, J. Theiss, L. Shi, and S. B. Cronin,
Direct observation of heat dissipation in individual suspended carbon nanotubes using a two-laser technique, \emph{J. Appl. Phys.}, \textbf{110}:044328, 2011.
\item A. I. Persson, Y. K. Koh, D. G. Cahill, L. Samuelson, and H. Linke, Thermal conductance of
InAs nanowire composites, \emph{Nano Lett.}, \textbf{9}:4484-4488, 2009.
\item F. Zhou, A. L. Moore, J. Bolinsson, A. Persson, L. Froberg, M. T. Pettes, H. Kong, L. Rabenberg, P. Caroff, D. A. Stewart, N. Mingo, K. A. Dick, L. Samuelson, H. Linke, and L. Shi,
Thermal conductivity of indium arsenide nanowires with wurtzite and zinc blende phases, \emph{Phys.
Rev. B}, \textbf{83}:205416, 2011.
\item J. Chung, K. Kim, G. Hwang, O. Kwon, S. Jung, J. Lee, J. W. Lee, and G. T. Kim, Quantitative temperature measurement of an electrically heated carbon nanotube using the null-point
method, \emph{Rev. Sci. Instru.}, \textbf{81}:114901, 2010.
\item K. Kim, J. Chung, J. Won, O. Kwon, J. S. Lee, S. H. Park, and Y. K. Choi, Quantitative scanning
thermal microscopy using double scan technique, \emph{Appl. Phys. Lett.}, \textbf{93}:3, 2008.
\item I. Jo, I. K. Hsu, Y. J. Lee, M. M. Sadeghi, S. Kim, S. Cronin, E. Tutuc, S. K. Banerjee,
Z. Yao, and L. Shi, Low-frequency acoustic phonon temperature distribution in electrically
biased graphene, \emph{Nano Lett.}, \textbf{11}:85-90, 2011.
\item Y.-M. Lin and M. S. Dresselhaus, Determination of carrier density in Te-doped Bi nanowires,
\emph{Appl. Phys. Lett.}, \textbf{83}:3567-3569, 2003.
\item J. Small and P. Kim, Thermopower measurement of individual single walled carbon nanotubes,
\emph{Microscale Thermophys. Eng.}, \textbf{8}:1-5, 2004.
\item J. H. Seol, A. L. Moore, S. K. Saha, F. Zhou, L. Shi, Q. L. Ye, R. Scheffler, N. Mingo, and
T. Yamada, Measurement and analysis of thermopower and electrical conductivity of an indium
antimonide nanowire from a vapor-liquid-solid method, \emph{J. Appl. Phys.}, \textbf{101}:023706, 2007.
\end{enumerate}

\end{document}